\documentclass[aps,prl,amsmath,showpacs,preprintnumbers,twocolumn,amssymb,groupedaddress]{revtex4-1}

\usepackage{graphicx}
\usepackage{mathptmx, textcomp}
\usepackage[latin1]{inputenc}
\usepackage{tabularx}
\usepackage{siunitx}
\usepackage{color}
\definecolor{bl}{rgb}{0, .1, .6}
\definecolor{rd}{rgb}{1,0,.2}

\usepackage[colorlinks=true, citecolor = bl, linkcolor = bl, urlcolor=bl, pdfborder={0 0 0}]{hyperref}
\usepackage{natbib}
\usepackage{tabularx}
\usepackage{lipsum}
\newcommand{\mum}{\SI{}{\micro\meter}}

\newcommand{\Hz}{{\rm Hz}}

\newcommand{\be}{\begin{eqnarray}}
\newcommand{\ee}{\end{eqnarray}}

\newcommand{\edd}{\epsilon_{\rm dd}}

\begin{document}

\title{Liquid quantum droplets of ultracold magnetic atoms}

\author{Igor Ferrier-Barbut}
\author{Matthias Schmitt}
\author{Matthias Wenzel}
\author{Holger Kadau}
\author{Tilman Pfau}
\affiliation{5. Physikalisches Institut and Center for Integrated Quantum Science and Technology,
Universit\"at Stuttgart, Pfaffenwaldring 57, 70550 Stuttgart, Germany}


\begin{abstract}
The simultaneous presence of two competing inter-particle interactions can lead to the emergence of new phenomena in a many-body system. Among others, such effects are expected in dipolar Bose-Einstein condensates, subject to dipole-dipole interaction and short-range repulsion. Magnetic quantum gases and in particular Dysprosium gases, offering a comparable short-range contact and a long-range dipolar interaction energy, remarkably exhibit such emergent phenomena. In addition an effective cancellation of mean-field effects of the two interactions results in a pronounced importance of quantum-mechanical beyond mean-field effects. For a weakly-dominant dipolar interaction the striking consequence is the existence of a new state of matter equilibrated by the balance between weak mean-field attraction and beyond mean-field repulsion. Though exemplified here in the case of dipolar Bose gases, this state of matter should appear also with other microscopic interactions types, provided a competition results in an effective cancellation of the total mean-field. The macroscopic state takes the form of so-called quantum droplets. We present the effects of a long-range dipolar interaction between these droplets. 
\end{abstract}

\maketitle

\section{Introduction}

The field of dipolar gases has emerged more than ten years ago, driven by the search for novel quantum many-body phenomena that would arise from the peculiarity of the dipole-dipole interaction: its long-range and anisotropic character \cite{Griesmaier:2005,Beaufils:2008}. This advent has lead to the observation of a large span of dipolar effects at the many-body level ranging from magnetostriction to collective excitations anisotropy or demagnetization cooling, among others \cite{Stuhler:2005,Fattori:2006,Lahaye:2007,Koch:2008,Lahaye:2008,Bismut:2010,Pasquiou:2011,Pasquiou:2012,Bismut:2012,dePaz:2013,Aikawa:2014a,Aikawa:2014b,Burdick:2015,Baier201,Tang:2016}.\par
The theoretical treatment of harmonically trapped dipolar bosonic gases is greatly simplified by applying an ansatz for the atomic density distribution, assuming either a gaussian distribution or an inverted parabola, solution of the Thomas-Fermi approximation at the mean-field level. Though successful in predicting dipolar phenomena such as magnetostriction, mean-field stability threshold and low-lying excitation energies, fixing the density distribution freezes out phase and density modulations within the atomic cloud. Thus this effectively neglects effects related to excitations at finite wavelength. However, learning from classical ferrofluids \cite{Rosensweig:1985} we know that in a magnetic fluid, it is finite-wavelengths that lead to the Rosensweig or normal-field instability creating stable surface modulations, thus they should also not be neglected in a dipolar quantum system.\par


A dipolar Bose-Einstein condensate is subject to two interparticle interactions, the contact interaction $V_{\rm c}(\boldsymbol r)=\frac{4\pi \hbar^2a}{m}\delta(\boldsymbol r)$ with the scattering length $a$, and the dipole interaction $V_{\rm dd}=\frac{3\hbar^2a_{\rm dd}}{m}\frac{(1-3\cos^2\theta)}{r^3}$\footnote{While the model presented here, based on the two simple interaction potentials is sufficient to describe the experimental results observed so far, precise quantitative measurements might reveal new physics, in particular the interaction potential could take a more subtle form.} with the dipole length $a_{\rm dd}=\frac{\mu_0\mu^2m}{12\pi\hbar^2}$ from which $\varepsilon_{\rm dd}=a_{\rm dd}/a$ is defined. In addition to this, the atoms are typically subject to an external harmonic trapping potential. The first indication of the important role of finite wavelength instability came with the prediction of a `roton' minimum in the dispersion relation of the excitations of a flattened dipolar Bose gas \cite{Santos:2003}. A wealth of theoretical studies suggested possible experimental evidences for such roton modes \cite{Wilson:2008,Lahaye:2009,Wilson:2009a,Wilson:2009b,Nath:2010,Ticknor:2011,Jona-Lasinio:2013,Bisset:2013a,Bisset:2013b}, with in particular a prediction of an instability of `angular roton' modes in a trapped flattened dipolar Bose gas \cite{Ronen:2007,Wilson:2009a}. Such instability stems directly from the long-range and anisotropic character of the DDI. 

\section{Experimental Observations}
\begin{figure*}[hbt]
\begin{center}
\begin{tabular}{cc}
  \hspace{-1cm}\includegraphics[width=.7\textwidth]{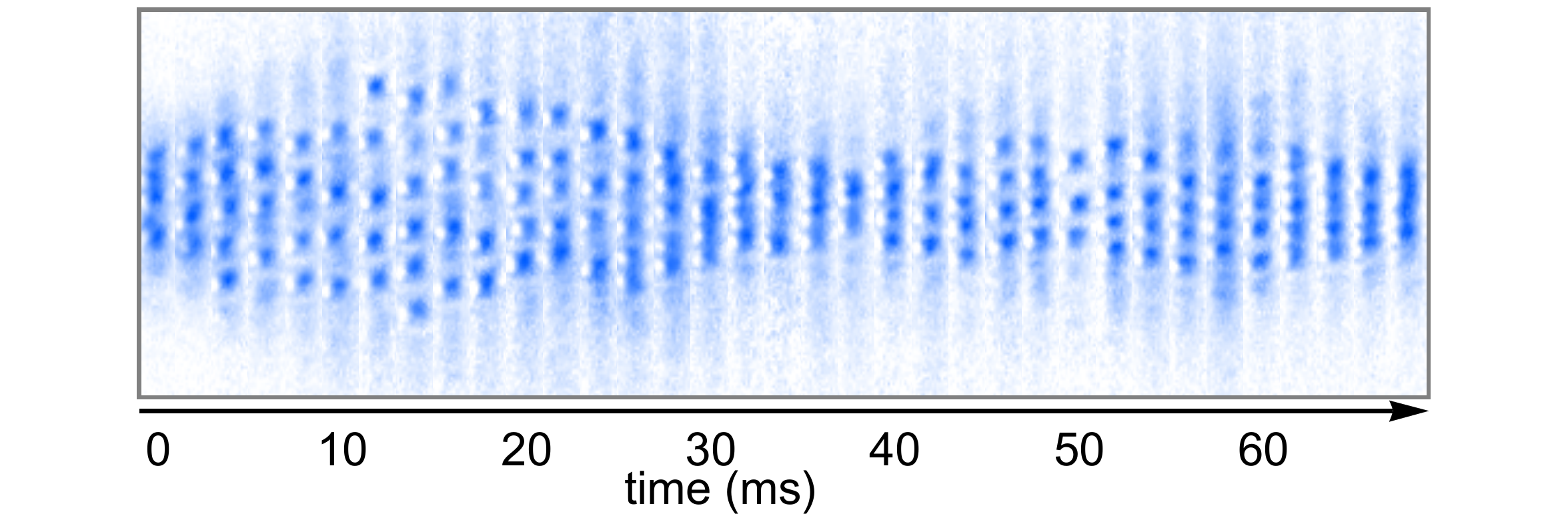}&
  \hspace{-1cm}\includegraphics[width=.35\textwidth]{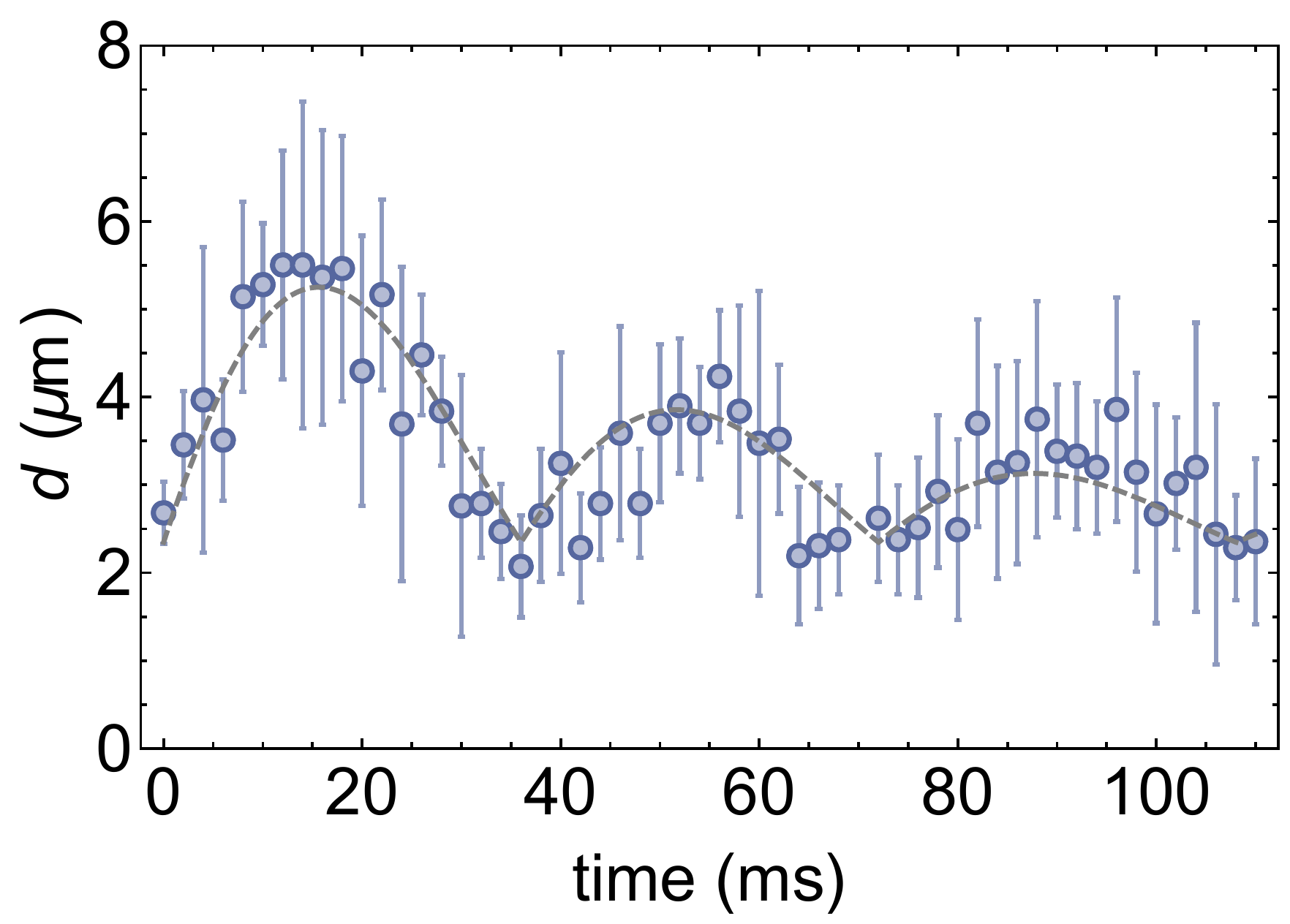}
\end{tabular}
\end{center}
  \caption{Experimental observation of droplet collisions in a an elongated trap. Left: single-shot images of droplets oscillating in the elongated trap at different times following a suddent transfer from a tight crossed trap (at $t=0$). The center of-mass motion has been subtracted in these images to display clearly the relative motion of the droplets. The magnetic field axis is pointing out of the imaging plane, thus the droplet elongation along this axis is not visible. Right: mean separation $d$ between neighbour droplets as a function of time, the error bars represent one standard deviation with 5 realisations per point, the gray dashed line is a guide to the eye representing a damped bouncing motion at the trap frequency $\nu_x=14.5\,\Hz$.}
  \label{Fig:Bounce}
\end{figure*}
While the prediction for this instability, which differs in nature from the one occurring in attractive Bose gases, has been intensively studied theoretically, until recently it eluded experimental observation. This instability occurs for lowering the scattering length (increasing $\varepsilon_{\rm dd}$), thus requiring interaction control through Feshbach resonances (F.R.) \cite{Chin:2008}. The observation of this instability was mostly prevented by the very short lifetime of chromium gases in the strongly dipolar regime ($\varepsilon_{\rm dd}\geqslant1$). This short lifetime of Cr gases results from the fact that the condition $\edd>1$ is only fulfilled in the close vicinity of the zero crossing ($a=0$) of a F.R. where three-body losses are enhanced \cite{Lahaye:2007}. Degenerate quantum gases of magnetic lanthanide atoms dysprosium and erbium, which have been obtained in recent years \cite{Lu:2011,Tang:2015b,Aikawa:2012} lifted this difficulty by offering a gain of a factor $5-10$ in $a_{\rm dd}$ thus allowing to work farther away from zero crossings. In addition, the typical wavelength at which this instability occurs is of order $1\mum$ thus the need for a high spatial resolution imaging. Our team in Stuttgart has developed a setup allowing for an in-situ imaging resolution of one micron. Whereas in the case of Cr where the relative simplicity of the atomic structure limits the number of F.R. and allows for their understanding \cite{Werner:2005}, the sub-merged open $4f$ shell of Lanthanide atoms creates a dense, strongly-correlated distribution of resonances reminiscent of quantum chaos \cite{Frisch:2014,Maier:2015a}, beyond the reach of ab initio understanding. Experimental efforts have thus been devoted to the understanding of scattering properties and mapping of the Feshbach spectrum of Dy and Er, \cite{Frisch:2014,Baumann:2014,Maier:2015a,Maier:2015b,Frisch:2015,Tang:2015a,Tang:2016}, rendering possible interaction control for many-body studies.\par
We thus have been able to study the behaviour of BECs in the vicinity of the transition between the short-range dominated ($\varepsilon_{\rm dd}<1$) and dipole dominated  ($\varepsilon_{\rm dd}>1$) regimes. To do so we have placed the BEC in an oblate trap (pancake configuration) with cylindrical symmetry around the dipoles direction (aligned by the magnetic field $\boldsymbol B\,\parallel\,\boldsymbol z$). Our observations have revealed the existence of an instability of a trapped BEC, located close to $\varepsilon_{\rm dd}=1$. They are performed on the positive $a$-tail of a Feshbach resonance with \textsuperscript{164}Dy, namely we are able to tune the scattering length around the background value $a_{\rm bg}=92(8)\,a_0$ \cite{Tang:2015a,Maier:2015b}. However this was measured in specific magnetic field ranges, and in the vicinity of a particular Feshbach resonance this value might differ, future precision measurements are still needed. Nevertheless, a relative tunability of $\varepsilon_{\rm dd}$ in the vicinity of $\edd=1$ is experimentally available, far from this point ($\edd\gg1$, $\edd\ll1$) Feshbach-enhanced three-body losses hamper the lifetime of the atomic ensemble on a narrow Feshbach resonance.\par
Using our high-resolution in-situ imaging, we have observed the time evolution of the integrated density distribution after a quench of $\varepsilon_{\rm dd}$. Following the quench, we observed the splitting of the density distribution into multiple density peaks \cite{Kadau:2016a}. The finite-wavelength nature of the instability was made clear by the appearance of a peak in Fourier space, that remains after azimuthal averaging. This character is directly due to the characteristics of the DDI, as was predicted. The exact mechanism triggering the instability remains to be determined, in particular the role of fluctuations (thermal and quantum) should be explored, they are likely to be this trigger. The very existence of this finite-wavelength instability is a major benchmark of many-body dipolar physics, however a very interesting surprise was discovered in its product. Indeed, the common wisdom in the field prior to our observations, based on mean-field calculations, was that this instability would be followed by a subsequent collapse of the individual density peaks. In very stark contrast to this prediction, the gas forms droplet ensembles that decay on very long time scales, on the order of several tens to hundreds of ms. \par
\begin{figure*}[hbt]
\begin{center}
\includegraphics[width=\textwidth]{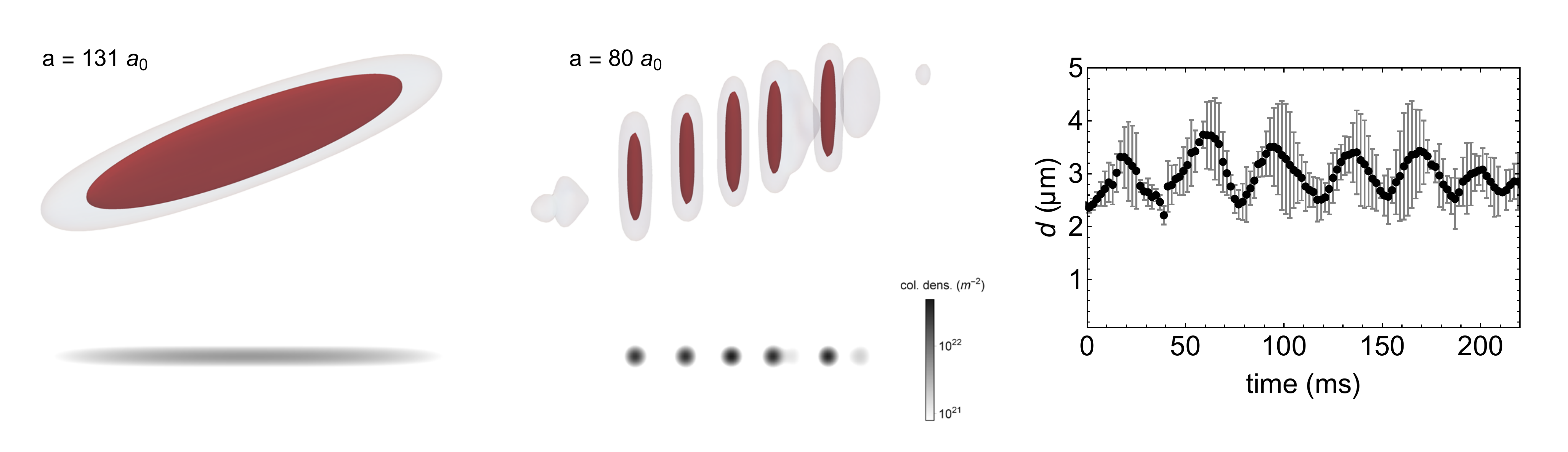}
\end{center}
  \caption{Effective Gross-Pitaevskii equation (\ref{Eq:EffectiveGPE}) simulations result. Left and central panel show isodensity surfaces at $20\%$ (red) and $1\%$ of the maximal density, the bottom images show integrated densities along the field direction. Left: at high scattering length ($\edd=1$), a BEC  is obtained with max density: $2.2\times10^{20}\,{\rm m}^{-3}$, while in the center for $\edd=1.64$ the simulations result in the formation of droplets with maximal density $2.1\times10^{21}\,{\rm m}^{-3}$, similarly to the experimental results of figure\,\ref{Fig:Bounce}. On the right one sees the time evolution of the mean next neighbour spacing between the 5 droplets as a function of time, oscillating in a trap a the same frequency as the experimental one $\nu_x=14.5\,\Hz$, showing a very similar behaviour as in figure\,\ref{Fig:Bounce}. The data on the right panel have been obtained by a single simulation, the points are the mean spacing between the five droplets obtained and the error bars show the standard deviation in the spacings.}
  \label{Fig:BounceSim}
\end{figure*}
The long-range repulsive nature of the interaction between the droplets is evident in images where we observe the formation of quasi-triangular lattice structure at the level of a few droplets. In order to verify that the confinement of these droplets is not due to the repulsive interaction with neighbouring droplets, and that the mechanism ensuring their stability takes place locally within individual droplets, we have placed them into a waveguide perpendicular to the dipoles direction, without confinement along the waveguide's axis \cite{Ferrier-Barbut:2016}. In this situation, the inter-droplet repulsion causes them to fly away from each other, remaining both confined and stable. An interesting result was obtained by adding a weak harmonic confinement along the waveguide's axis, forcing the droplets to collide back on each other. In this case, although the droplet ensemble creation is stochastic in nature, resulting in different experimental conditions (number of droplets, exact initial inter-droplet separation), we observe a clear bouncing motion of the droplets on each other. This is represented in figure\,\ref{Fig:Bounce}, where one observes that the droplets bounce off each other twice in the waveguide, as a result of the dipolar repulsion that exists between them. In these experiments we first created droplet ensembles in a crossed trap in which they were tightly confined in all three directions, before releasing them in a very weak elongated trap along the $x$ axis with weak frequency $\nu_x=14.5\,\Hz$. This results in a high initial amplitude of oscillation due to the initial repulsion in the crossed trap. Using a simple model considering the droplets as point-like dipoles linearized around the equilibrium solution one can easily show that the expected frequency for small oscillations is $\nu_x\,\sqrt{l+2}$ for a repulsive potential with power law $1/r^l$\footnote{The next order (nonlinearity) grows like $(l+2)(l+3)$}, thus a precise frequency measurement at very low oscillation amplitude would be an excellent test of dipolar repulsion between droplets (for which $l=3$). Our measurements are performed at a too large amplitude to observe this and thus show a frequency close to $2\nu_x$ which is expected for these amplitudes. They also exhibit a strong damping. The mechanism for this damping remains to be understood, it might be in part due to friction with a remnant thermal fraction (this fraction is hard to tell apart from BEC atoms not confined in droplets), but could also originate from inelastic collisions between droplets. At equilibrium obtained by adiabatically transferring the droplets into the weakly-trapping waveguide we measure an equilibrium distance between the droplets $d_{\rm eq}=2.5(5)\,\mum$. While point-like dipoles with dipole moment $\mu_N=N\mu$ and mass $m_N=Nm$ would equilibrate at $d=(3\mu_0\mu_{\rm N}^2/2\pi m_N\omega^2)^{1/5}\approx4.5\,\mum$, the elongation of the droplets reduces their effective moment to lower values than $N\,\mu$ where $N$ is the number of atoms they contain. Knowing $N$, we could obtain an upper bound on the droplet aspect ratio defined as $\kappa=R_r/R_z$: $\kappa\lesssim0.2$ with $R_r$ ($R_z$) the droplets radius perpendicular (parallel) to the field direction. While we have shown the existence of droplets and their interaction, we now turn to the mechanism responsible for their stability.\par

\section{Mean-field analysis}

Indeed, as a result of the anisotropy of the DDI and the droplets elongation along the magnetic field, the effective interactions at the mean-filed level are attractive. This can be seen easily by expressing the mean-field  energy density $e=E/V$ at the center of a droplet (cylindrical around the field direction, with central density $n_0$), neglecting the kinetic energy as well as the trapping potential \cite{Lahaye:2009,Lima:2012}.
\be
e_{\rm MF}(0)=\frac{gn_0^2}{2}(1-\varepsilon_{\rm dd}\,f(\kappa)),\label{Eq:MFE}
\ee
where $g=4\pi\hbar^2a/m$, and $f(\kappa)$ is a decreasing function of $\kappa$ with $f(\kappa \ll1)=1$, $f(\kappa \gg1)=-2$, $f(1)=0$, this result holds either for a gaussian or inverted-parabola density distribution in the droplet \cite{Lahaye:2009,ODell:2004,Eberlein:2005}. If $\edd<1$, the situation is stable independent of $\kappa$, energy is minimized by lowering density, which will eventually be confined by the trap (gas-like behaviour). For $\edd>1$ two situations can emerge; either $\kappa\gg1$, in which case the term in parenthesis is positive leading to the same behaviour as before, or $\kappa\ll1$ in which case energy is minimized by increasing more and more the density, leading to a collapse. One must note however that in this situation for $\edd\gtrsim1$, the two types of interactions effectively `screen' each other, reducing greatly the attraction with respect to a purely dipolar situation. This qualitative behaviour obtained by the simple equation (\ref{Eq:MFE}) is confirmed by a theoretical analysis including kinetic energy (quantum pressure) and trapping potential, either making a gaussian ansatz with which $\kappa$ can be self-consistently calculated from the trap aspect ratio $\lambda=\omega_z/\omega_r$, or using Gross-Pitaevskii simulations. The exact value of $\lambda$ for which the system becomes unstable at a given $\edd$ can be thus obtained. However the conclusion remains: at $\edd>1$, for the experimental conditions $\kappa\lesssim0.2$, the droplets should undergo collapse in mean-field theory. In the next section we describe the advances that have been made to explain this stability. 

\section{Beyond mean field theory}
\begin{figure*}[hbt]
\begin{center}
\begin{tabular}{cc}
\includegraphics[width=.5\textwidth]{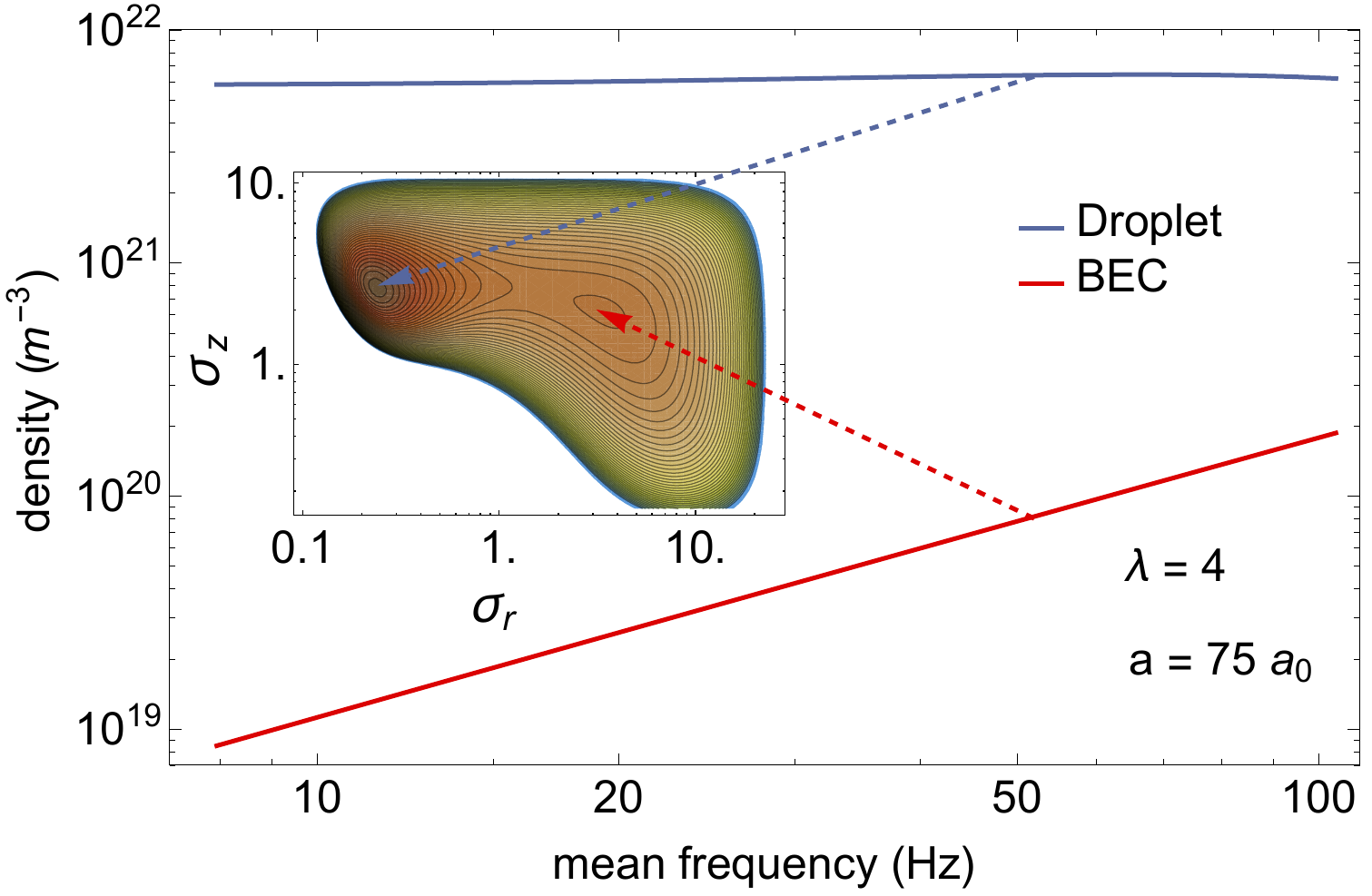}&
\includegraphics[width=.5\textwidth]{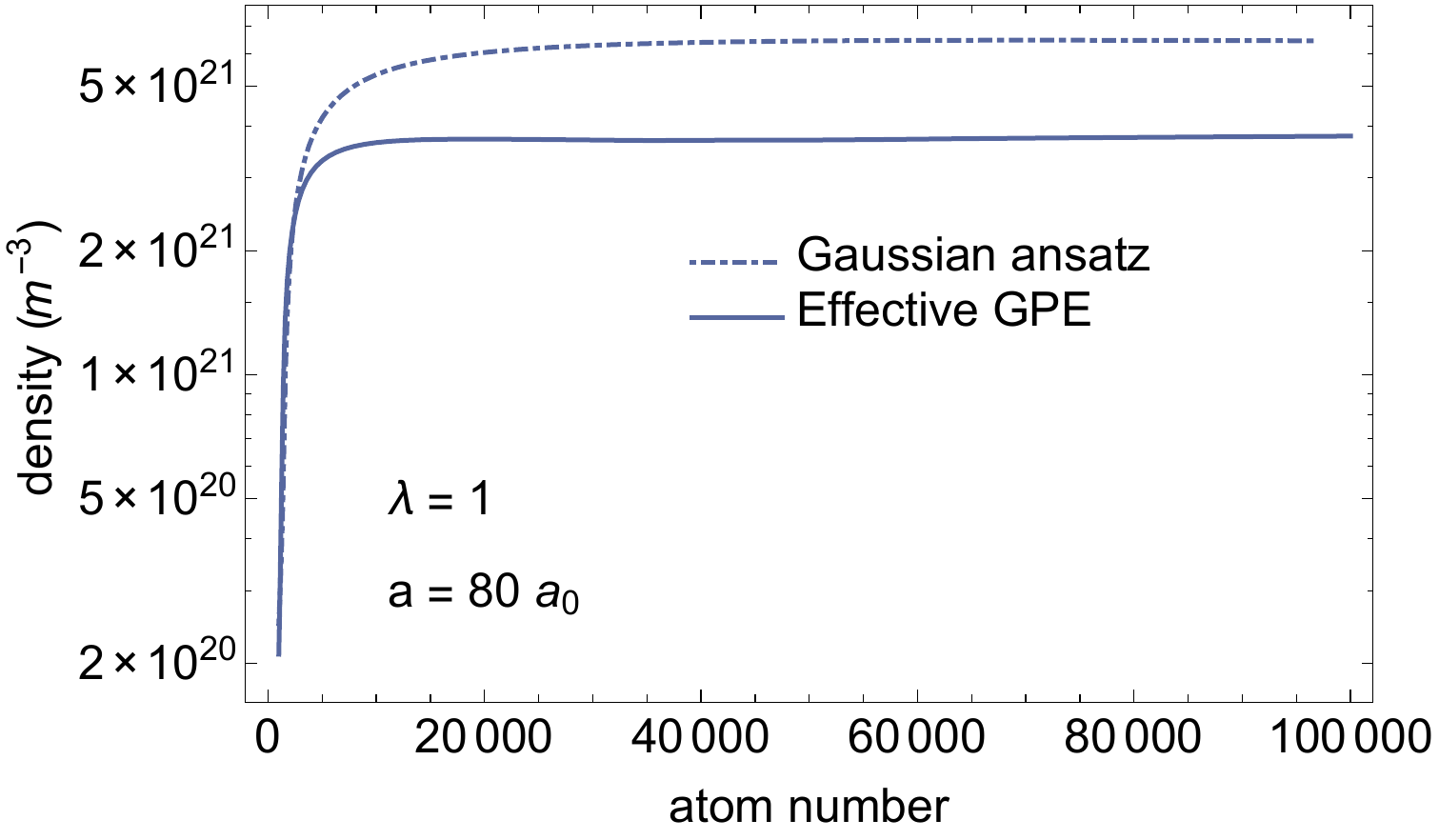}
\end{tabular}
\end{center}
  \caption{Important features of the liquid-like droplet phase. Left: In the region of the phase diagram where both the BEC and droplet phase coexist, central density using the gaussian ansatz as a function of mean frequency $\bar\omega/2\pi$ for a trap aspect ratio $\lambda=4$, scattering length $a=75\,a_0$ and atom number $N=5000$. The inset shows a typical energy landscape at $\bar\omega/2\pi=50\,\Hz$ as a function of radial and axial gaussian sizes in microns, from which two minima emerge. The striking difference between these two phases is clearly visible, while the BEC density scales as $\bar\omega^{6/5}$ as usual, the droplet density is nearly insensitive to the weak compression, behaving like a liquid. Right: Droplet density in an isotropic trap with $\omega/2\pi=100\,\Hz$, showing that the density reaches a plateau and becomes independent of the atom number. Once kinetic energy effects can be neglected, the effective GPE results display a lower density as the gaussian ansatz.}\label{Fig:Density}
\end{figure*}

Prior to our observations, theoretical proposals suggested different possible stabilization mechanisms for a collapsing BEC. One was three-body repulsion \cite{Bulgac:2002}. Put simply, this amounts to adding a mean-field term $\propto n^3$ in equation (\ref{Eq:MFE}), originating from a three-body interaction. While easy to treat theoretically, the appearance of this term at a significant magnitude is hard to justify from a microscopic model of interactions. Early theoretical analysis of the stable droplets came up with three-body repulsion as a stabilization mechanism \cite{Xi:2015,Bisset:2015,Blakie:2016}. However another proposed stabilization mechanism was beyond mean-field corrections \cite{Petrov:2015,Volovik:2009}. In contrast to the former, this is not added by hand but emerges naturally from quantum fluctuations/correlations, and in any case the effects cannot be neglected \cite{Ferrier-Barbut:2016}.  Several different groups, including the authors of \cite{Xi:2015,Bisset:2015,Blakie:2016} have now converged towards quantum fluctuations as a sufficient stabilization mechanism. To understand why, we will go on with our simplified model that neglects both trapping potential and kinetic energy, using the local density approximation (LDA). In this case the correction to the mean-field energy has been shown to be:
\be
\delta e_{\rm BMF}(0)=\frac{gn_0^2}{2}\;\frac{128\sqrt{n_0a^3}}{15\sqrt\pi}\,Q_5(\edd)\label{Eq:BMFE}
\ee
where the fraction $\frac{128\sqrt{n_0a^3}}{15\sqrt\pi}$ is the celebrated Lee-Huang-Yang result \cite{Lee:1957a,Lee:1957b} and the additional factor $Q_5(\edd)$ was obtained by Lima and Pelster in \cite{Lima:2011,Lima:2012}, following another work \cite{Scutzhold:2006}. This function $Q_5$ is in general complex, however in the experimentally relevant conditions $\edd\gtrsim1$, the imaginary part is very weak and we neglect it in the following. However one needs to keep in mind that it can play a role, notably by significantly destabilizing the system when $\edd\gg1$. In the regime of interest, it is also usually sufficient to take the lowest order expansion of $Q_5(\edd)\approx1+\frac32\edd^2$. We finally end up with a simple expression for the energy density:
\be
e(0)=\frac{gn_0^2}{2}(\alpha+\beta\sqrt{n_0}),\label{Eq:TotE}
\ee
with $\alpha=1-\edd f(\kappa)$ and $\beta=\frac{128\sqrt{a^3}(1+\frac32\edd^2)}{15\sqrt\pi}$, note that $\beta>0\, \forall\, a>0$.\par

In expression (\ref{Eq:TotE}), it is evident now that when $\alpha<0$, instead of collapsing, the density is stabilized by the additional term that has a stronger dependence on $n_0$ (liquid-like state). One can easily workout that the saturation density is $n_0\propto\left(\alpha/\beta\right)^2$, this tendency was confirmed in \cite{Ferrier-Barbut:2016}. This shows that different factors work at unison to reduce the liquid-like phase density, first the screening of the two interactions that significantly lowers $\alpha$, second the amplification of $\beta$ by the dipolar interaction. This density reduction is absolutely crucial because as it is most often the case in the least dilute regime of ultracold atoms, the lifetime is limited by density-dependent three-body losses ($\propto\langle n^2\rangle$). Expression (\ref{Eq:TotE}) encompasses the main ingredients explaining the existence of the droplet phase, but of course quantitative agreement can be obtained only when taking into account contributions from the quantum pressure and the trapping potexntial. In particular if one wants to describe both the finite-wavelength collapse and the droplets stabilization, this model is insufficient.\par
Usually, all these factors are taken into account performing numerical simulations of the Gross-Pitaevskii equation. Here, the additional term is beyond the GPE description, in fact quantum fluctuations deplete the ground state (the quantum depletion), by putting atoms in non-factorizable entangled states, which in principle cannot be described by a single field classical field theory (the GPE). However this quantum depletion remains very small in our regime \cite{Wachtler:2016a}, as a consequence, an effective classical field theory can be built by adding a term that reproduces (\ref{Eq:BMFE}). This has been employed in \cite{Wachtler:2016a}, and later-on in \cite{Wachtler:2016b,Bisset:2016}, and reproduces all the main features of our experiments. An important verification of this method was performed in \cite{Saito:2016} using path-integral quantum Monte Carlo, confirming its validity. Here, we perform  such simulations that reveal the most important properties of the droplets and droplet ensembles. The effective GPE reads:\par
\begin{small}
\begin{equation}
\begin{split}	
i\hbar\partial_t\psi(\boldsymbol{r})=\left[-\frac{\hbar^2\nabla^2}{2m}+V_{\rm ext}(\boldsymbol{r})+g|\psi(\boldsymbol{r})|^2+\right.\\
\left.\int d\boldsymbol{r'}|\psi(\boldsymbol{r'})|^2V_{\rm dd}(\boldsymbol{r}-\boldsymbol{r'})+\frac{32 \,g\,\sqrt{a^3}}{3\sqrt{\pi}}Q_5(\edd)|\psi(\boldsymbol r)|^3\right]\psi(\boldsymbol{r}),\label{Eq:EffectiveGPE}
\end{split}
\end{equation}
\end{small}
where the last term accounts for quantum fluctuations and is taken in the local density approximation. The time evolution of the wave function under eq.\,(\ref{Eq:EffectiveGPE}) is obtained using a usual split-step algorithm \cite{Wachtler:2016a,Bisset:2016}. We study here the evolution of a BEC first at rest in the elongated trap, then applying a quench in scattering length from $a=131\,a_0$ down to $a=80\,a_0$. This differs from the experimental procedure where the droplets are formed in a tight trap before being released in the elongated trap. The results from the simulations are presented in figure\,\ref{Fig:BounceSim}. We observe the formation of an array of droplets, very similar to the experimental observations presented earlier. In addition, the simulations confirm the bouncing motion of the droplets in the weakly trapping waveguide, where we have used the experimental trapping frequency along the WG direction. Thus the main features of the droplets array in the WG are very well reproduced. The much lower amplitude with respect to experiments (fig.\,\ref{Fig:Bounce}) is explained by the fact that the simulations start with a BEC initially at rest. To confirm the peculiar character of this novel state, we show below that it differs from a normal BEC.\par

For this purpose, we study theoretically the properties of a single droplet placed in a harmonic trap. First, it is instructive to apply a variational gaussian ansatz for the wave function, allowing to minimize the energy functional corresponding to (\ref{Eq:EffectiveGPE}) \cite{Kadau:2016b}. One finds that in a trap, two energy minima can exist, this can be easily understood by looking at expression (\ref{Eq:TotE}), indeed the first term allows for two situations, $\kappa\gg1$, and thus $\alpha>0$, the BEC is stabilized by dipolar repulsion, and behaves like a gas in the sense that minimizing energy is done by minimizing density. This first case can be realized only if the trapping potential forces a pancake-like shape to the cloud, thus only for high $\lambda$. Second, for $\kappa\ll1$ thus $\alpha<0$, the cloud will be stabilized by quantum fluctuations, at a fixed density, thus behaving more like a liquid. This has been studied in details in \cite{Wachtler:2016a,Wachtler:2016b,Bisset:2016}. To compare the two cases, we study in figure\,\ref{Fig:Density} how the central density evolves when $(i)$ the trap is compressed (left) and $(ii)$ when the atom number is increased (right). For $(i)$ we keep the trap aspect ratio constant, increasing the mean frequency $\bar\omega=(\omega_\rho^2\omega_z)$. We find a stark difference in behaviour: while the gas-like phase behaves as expected, with a density increasing with $\bar \omega$, we obtain that the liquid-like phase density is almost insensitive to the trap compression. In the case $(ii)$, that we study in an isotropic trap ($\lambda=1$), at low atom number for the gaussian ansatz the density depends on $N$ due to the action of quantum pressure, but then reaches a plateau that depends only very weakly on $N$ as expected for a liquid-like phase. For $\lambda=1$, the gas-like phase is not a solution of the gaussian ansatz for the parameters used here. Finally we perform a study of the ground state by imaginary-time evolution of (\ref{Eq:EffectiveGPE}), which allows to confirm the qualitative results obtained by the variational ansatz. Importantly, we find that the variational ansatz method overestimates the central density with respect to simulation results. This can be understood by the fact that interactions flatten the density distribution with respect to a gaussian ansatz \cite{Wachtler:2016a,Wachtler:2016b,Bisset:2016}.\par
The properties outlined here mark a strong difference with all other systems observed in quantum gases, since they are characteristic of an essentially incompressible liquid rather than a gas. Indeed the very simple model (\ref{Eq:TotE}) possesses a gas to liquid phase transition at $\alpha=0$, characterized by a divergence of the compressibility. A liquid differs from a gas in that when placed in a given volume, it does not fill the whole volume but remains self-bound. This self-bound liquid is observed in the absence of trapping \cite{Schmitt:2016}, as is expected \cite{Wachtler:2016b,Baillie:2016}. While these droplets have been discovered in dipolar Bose gases as recently confirmed with Erbium in the trapped case \cite{Chomaz:2016}, they are a manifestation of a generic phase that emerges at low density when two competing interactions are present. They should be observable for instance in mixtures of contact-interacting bosons \cite{Petrov:2015}. Thus a new field of exploration is opening in ultracold gases. The physics of these liquid droplets can be put to test by measurement of collective excitation frequencies, their superfluidity could be verified by the observation of trapped vortices inside a single droplet. These observation could allow to bridge this field with that of liquid helium droplets \cite{Dalfovo:2001}, for which such collective modes have been calculated, and vortices have been observed \cite{Casas:1990,Gomez:2014}. Experimentally, a trapping potential is no longer necessary, which will lead to new possibilities, for instance in terms of transport which can be realised with a simple magnetic gradient. 
\begin{acknowledgements}
	We thank Luis Santos and Francesca Ferlaino for discussions, and acknowledge financial support from the Deutsche Forschungsgemeinschaft through SFB/TRR 21 and the Forschergruppe FOR 2247, I.F.B. acknowledges support from the EU within a Horizon2020 Marie Sk\l odowska-Curie IF (703419 DipInQuantum).  
\end{acknowledgements}

\bibliographystyle{apsrev4-1}

\begin{thebibliography}{0}%
\makeatletter
\providecommand \@ifxundefined [1]{%
 \@ifx{#1\undefined}
}%
\providecommand \@ifnum [1]{%
 \ifnum #1\expandafter \@firstoftwo
 \else \expandafter \@secondoftwo
 \fi
}%
\providecommand \@ifx [1]{%
 \ifx #1\expandafter \@firstoftwo
 \else \expandafter \@secondoftwo
 \fi
}%
\providecommand \natexlab [1]{#1}%
\providecommand \enquote  [1]{``#1''}%
\providecommand \bibnamefont  [1]{#1}%
\providecommand \bibfnamefont [1]{#1}%
\providecommand \citenamefont [1]{#1}%
\providecommand \href@noop [0]{\@secondoftwo}%
\providecommand \href [0]{\begingroup \@sanitize@url \@href}%
\providecommand \@href[1]{\@@startlink{#1}\@@href}%
\providecommand \@@href[1]{\endgroup#1\@@endlink}%
\providecommand \@sanitize@url [0]{\catcode `\\12\catcode `\$12\catcode
  `\&12\catcode `\#12\catcode `\^12\catcode `\_12\catcode `\%12\relax}%
\providecommand \@@startlink[1]{}%
\providecommand \@@endlink[0]{}%
\providecommand \url  [0]{\begingroup\@sanitize@url \@url }%
\providecommand \@url [1]{\endgroup\@href {#1}{\urlprefix }}%
\providecommand \urlprefix  [0]{URL }%
\providecommand \Eprint [0]{\href }%
\providecommand \doibase [0]{http://dx.doi.org/}%
\providecommand \selectlanguage [0]{\@gobble}%
\providecommand \bibinfo  [0]{\@secondoftwo}%
\providecommand \bibfield  [0]{\@secondoftwo}%
\providecommand \translation [1]{[#1]}%
\providecommand \BibitemOpen [0]{}%
\providecommand \bibitemStop [0]{}%
\providecommand \bibitemNoStop [0]{.\EOS\space}%
\providecommand \EOS [0]{\spacefactor3000\relax}%
\providecommand \BibitemShut  [1]{\csname bibitem#1\endcsname}%
\let\auto@bib@innerbib\@empty
\end{thebibliography}%


\begin{thebibliography}{68}%
\makeatletter
\providecommand \@ifxundefined [1]{%
 \@ifx{#1\undefined}
}%
\providecommand \@ifnum [1]{%
 \ifnum #1\expandafter \@firstoftwo
 \else \expandafter \@secondoftwo
 \fi
}%
\providecommand \@ifx [1]{%
 \ifx #1\expandafter \@firstoftwo
 \else \expandafter \@secondoftwo
 \fi
}%
\providecommand \natexlab [1]{#1}%
\providecommand \enquote  [1]{``#1''}%
\providecommand \bibnamefont  [1]{#1}%
\providecommand \bibfnamefont [1]{#1}%
\providecommand \citenamefont [1]{#1}%
\providecommand \href@noop [0]{\@secondoftwo}%
\providecommand \href [0]{\begingroup \@sanitize@url \@href}%
\providecommand \@href[1]{\@@startlink{#1}\@@href}%
\providecommand \@@href[1]{\endgroup#1\@@endlink}%
\providecommand \@sanitize@url [0]{\catcode `\\12\catcode `\$12\catcode
  `\&12\catcode `\#12\catcode `\^12\catcode `\_12\catcode `\%12\relax}%
\providecommand \@@startlink[1]{}%
\providecommand \@@endlink[0]{}%
\providecommand \url  [0]{\begingroup\@sanitize@url \@url }%
\providecommand \@url [1]{\endgroup\@href {#1}{\urlprefix }}%
\providecommand \urlprefix  [0]{URL }%
\providecommand \Eprint [0]{\href }%
\providecommand \doibase [0]{http://dx.doi.org/}%
\providecommand \selectlanguage [0]{\@gobble}%
\providecommand \bibinfo  [0]{\@secondoftwo}%
\providecommand \bibfield  [0]{\@secondoftwo}%
\providecommand \translation [1]{[#1]}%
\providecommand \BibitemOpen [0]{}%
\providecommand \bibitemStop [0]{}%
\providecommand \bibitemNoStop [0]{.\EOS\space}%
\providecommand \EOS [0]{\spacefactor3000\relax}%
\providecommand \BibitemShut  [1]{\csname bibitem#1\endcsname}%
\let\auto@bib@innerbib\@empty
\bibitem [{\citenamefont {Griesmaier}\ \emph {et~al.}(2005)\citenamefont
  {Griesmaier}, \citenamefont {Werner}, \citenamefont {Hensler}, \citenamefont
  {Stuhler},\ and\ \citenamefont {Pfau}}]{Griesmaier:2005}%
  \BibitemOpen
  \bibfield  {author} {\bibinfo {author} {\bibfnamefont {A.}~\bibnamefont
  {Griesmaier}}, \bibinfo {author} {\bibfnamefont {J.}~\bibnamefont {Werner}},
  \bibinfo {author} {\bibfnamefont {S.}~\bibnamefont {Hensler}}, \bibinfo
  {author} {\bibfnamefont {J.}~\bibnamefont {Stuhler}}, \ and\ \bibinfo
  {author} {\bibfnamefont {T.}~\bibnamefont {Pfau}},\ }\href {\doibase
  10.1103/PhysRevLett.94.160401} {\bibfield  {journal} {\bibinfo  {journal}
  {Phys. Rev. Lett.}\ }\textbf {\bibinfo {volume} {94}},\ \bibinfo {pages}
  {160401} (\bibinfo {year} {2005})}\BibitemShut {NoStop}%
\bibitem [{\citenamefont {Beaufils}\ \emph {et~al.}(2008)\citenamefont
  {Beaufils}, \citenamefont {Chicireanu}, \citenamefont {Zanon}, \citenamefont
  {Laburthe-Tolra}, \citenamefont {Mar\'echal}, \citenamefont {Vernac},
  \citenamefont {Keller},\ and\ \citenamefont {Gorceix}}]{Beaufils:2008}%
  \BibitemOpen
  \bibfield  {author} {\bibinfo {author} {\bibfnamefont {Q.}~\bibnamefont
  {Beaufils}}, \bibinfo {author} {\bibfnamefont {R.}~\bibnamefont
  {Chicireanu}}, \bibinfo {author} {\bibfnamefont {T.}~\bibnamefont {Zanon}},
  \bibinfo {author} {\bibfnamefont {B.}~\bibnamefont {Laburthe-Tolra}},
  \bibinfo {author} {\bibfnamefont {E.}~\bibnamefont {Mar\'echal}}, \bibinfo
  {author} {\bibfnamefont {L.}~\bibnamefont {Vernac}}, \bibinfo {author}
  {\bibfnamefont {J.-C.}\ \bibnamefont {Keller}}, \ and\ \bibinfo {author}
  {\bibfnamefont {O.}~\bibnamefont {Gorceix}},\ }\href {\doibase
  10.1103/PhysRevA.77.061601} {\bibfield  {journal} {\bibinfo  {journal} {Phys.
  Rev. A}\ }\textbf {\bibinfo {volume} {77}},\ \bibinfo {pages} {061601}
  (\bibinfo {year} {2008})}\BibitemShut {NoStop}%
\bibitem [{\citenamefont {Stuhler}\ \emph {et~al.}(2005)\citenamefont
  {Stuhler}, \citenamefont {Griesmaier}, \citenamefont {Koch}, \citenamefont
  {Fattori}, \citenamefont {Pfau}, \citenamefont {Giovanazzi}, \citenamefont
  {Pedri},\ and\ \citenamefont {Santos}}]{Stuhler:2005}%
  \BibitemOpen
  \bibfield  {author} {\bibinfo {author} {\bibfnamefont {J.}~\bibnamefont
  {Stuhler}}, \bibinfo {author} {\bibfnamefont {A.}~\bibnamefont {Griesmaier}},
  \bibinfo {author} {\bibfnamefont {T.}~\bibnamefont {Koch}}, \bibinfo {author}
  {\bibfnamefont {M.}~\bibnamefont {Fattori}}, \bibinfo {author} {\bibfnamefont
  {T.}~\bibnamefont {Pfau}}, \bibinfo {author} {\bibfnamefont {S.}~\bibnamefont
  {Giovanazzi}}, \bibinfo {author} {\bibfnamefont {P.}~\bibnamefont {Pedri}}, \
  and\ \bibinfo {author} {\bibfnamefont {L.}~\bibnamefont {Santos}},\ }\href
  {\doibase 10.1103/PhysRevLett.95.150406} {\bibfield  {journal} {\bibinfo
  {journal} {Phys. Rev. Lett.}\ }\textbf {\bibinfo {volume} {95}},\ \bibinfo
  {pages} {150406} (\bibinfo {year} {2005})}\BibitemShut {NoStop}%
\bibitem [{\citenamefont {Fattori}\ \emph {et~al.}(2006)\citenamefont
  {Fattori}, \citenamefont {Koch}, \citenamefont {Goetz}, \citenamefont
  {Griesmaier}, \citenamefont {Hensler}, \citenamefont {Stuhler},\ and\
  \citenamefont {Pfau}}]{Fattori:2006}%
  \BibitemOpen
  \bibfield  {author} {\bibinfo {author} {\bibfnamefont {M.}~\bibnamefont
  {Fattori}}, \bibinfo {author} {\bibfnamefont {T.}~\bibnamefont {Koch}},
  \bibinfo {author} {\bibfnamefont {S.}~\bibnamefont {Goetz}}, \bibinfo
  {author} {\bibfnamefont {A.}~\bibnamefont {Griesmaier}}, \bibinfo {author}
  {\bibfnamefont {S.}~\bibnamefont {Hensler}}, \bibinfo {author} {\bibfnamefont
  {J.}~\bibnamefont {Stuhler}}, \ and\ \bibinfo {author} {\bibfnamefont
  {T.}~\bibnamefont {Pfau}},\ }\href {http://dx.doi.org/10.1038/nphys443}
  {\bibfield  {journal} {\bibinfo  {journal} {Nat Phys}\ }\textbf {\bibinfo
  {volume} {2}},\ \bibinfo {pages} {765} (\bibinfo {year} {2006})}\BibitemShut
  {NoStop}%
\bibitem [{\citenamefont {Lahaye}\ \emph {et~al.}(2007)\citenamefont {Lahaye},
  \citenamefont {Koch}, \citenamefont {Fr{\"o}hlich}, \citenamefont {Fattori},
  \citenamefont {Metz}, \citenamefont {Griesmaier}, \citenamefont
  {Giovanazzi},\ and\ \citenamefont {Pfau}}]{Lahaye:2007}%
  \BibitemOpen
  \bibfield  {author} {\bibinfo {author} {\bibfnamefont {T.}~\bibnamefont
  {Lahaye}}, \bibinfo {author} {\bibfnamefont {T.}~\bibnamefont {Koch}},
  \bibinfo {author} {\bibfnamefont {B.}~\bibnamefont {Fr{\"o}hlich}}, \bibinfo
  {author} {\bibfnamefont {M.}~\bibnamefont {Fattori}}, \bibinfo {author}
  {\bibfnamefont {J.}~\bibnamefont {Metz}}, \bibinfo {author} {\bibfnamefont
  {A.}~\bibnamefont {Griesmaier}}, \bibinfo {author} {\bibfnamefont
  {S.}~\bibnamefont {Giovanazzi}}, \ and\ \bibinfo {author} {\bibfnamefont
  {T.}~\bibnamefont {Pfau}},\ }\href {\doibase 10.1038/nature06036} {\bibfield
  {journal} {\bibinfo  {journal} {Nature}\ }\textbf {\bibinfo {volume} {448}},\
  \bibinfo {pages} {672} (\bibinfo {year} {2007})}\BibitemShut {NoStop}%
\bibitem [{\citenamefont {Koch}\ \emph {et~al.}(2008)\citenamefont {Koch},
  \citenamefont {Lahaye}, \citenamefont {Metz}, \citenamefont {Fr{\"o}hlich},
  \citenamefont {Griesmaier},\ and\ \citenamefont {Pfau}}]{Koch:2008}%
  \BibitemOpen
  \bibfield  {author} {\bibinfo {author} {\bibfnamefont {T.}~\bibnamefont
  {Koch}}, \bibinfo {author} {\bibfnamefont {T.}~\bibnamefont {Lahaye}},
  \bibinfo {author} {\bibfnamefont {J.}~\bibnamefont {Metz}}, \bibinfo {author}
  {\bibfnamefont {B.}~\bibnamefont {Fr{\"o}hlich}}, \bibinfo {author}
  {\bibfnamefont {A.}~\bibnamefont {Griesmaier}}, \ and\ \bibinfo {author}
  {\bibfnamefont {T.}~\bibnamefont {Pfau}},\ }\href {\doibase
  doi:10.1038/nphys887} {\bibfield  {journal} {\bibinfo  {journal} {Nature
  Physics}\ }\textbf {\bibinfo {volume} {4}},\ \bibinfo {pages} {218} (\bibinfo
  {year} {2008})}\BibitemShut {NoStop}%
\bibitem [{\citenamefont {Lahaye}\ \emph {et~al.}(2008)\citenamefont {Lahaye},
  \citenamefont {Metz}, \citenamefont {Fr\"ohlich}, \citenamefont {Koch},
  \citenamefont {Meister}, \citenamefont {Griesmaier}, \citenamefont {Pfau},
  \citenamefont {Saito}, \citenamefont {Kawaguchi},\ and\ \citenamefont
  {Ueda}}]{Lahaye:2008}%
  \BibitemOpen
  \bibfield  {author} {\bibinfo {author} {\bibfnamefont {T.}~\bibnamefont
  {Lahaye}}, \bibinfo {author} {\bibfnamefont {J.}~\bibnamefont {Metz}},
  \bibinfo {author} {\bibfnamefont {B.}~\bibnamefont {Fr\"ohlich}}, \bibinfo
  {author} {\bibfnamefont {T.}~\bibnamefont {Koch}}, \bibinfo {author}
  {\bibfnamefont {M.}~\bibnamefont {Meister}}, \bibinfo {author} {\bibfnamefont
  {A.}~\bibnamefont {Griesmaier}}, \bibinfo {author} {\bibfnamefont
  {T.}~\bibnamefont {Pfau}}, \bibinfo {author} {\bibfnamefont {H.}~\bibnamefont
  {Saito}}, \bibinfo {author} {\bibfnamefont {Y.}~\bibnamefont {Kawaguchi}}, \
  and\ \bibinfo {author} {\bibfnamefont {M.}~\bibnamefont {Ueda}},\ }\href
  {\doibase 10.1103/PhysRevLett.101.080401} {\bibfield  {journal} {\bibinfo
  {journal} {Phys. Rev. Lett.}\ }\textbf {\bibinfo {volume} {101}},\ \bibinfo
  {pages} {080401} (\bibinfo {year} {2008})}\BibitemShut {NoStop}%
\bibitem [{\citenamefont {Bismut}\ \emph {et~al.}(2010)\citenamefont {Bismut},
  \citenamefont {Pasquiou}, \citenamefont {Mar\'echal}, \citenamefont {Pedri},
  \citenamefont {Vernac}, \citenamefont {Gorceix},\ and\ \citenamefont
  {Laburthe-Tolra}}]{Bismut:2010}%
  \BibitemOpen
  \bibfield  {author} {\bibinfo {author} {\bibfnamefont {G.}~\bibnamefont
  {Bismut}}, \bibinfo {author} {\bibfnamefont {B.}~\bibnamefont {Pasquiou}},
  \bibinfo {author} {\bibfnamefont {E.}~\bibnamefont {Mar\'echal}}, \bibinfo
  {author} {\bibfnamefont {P.}~\bibnamefont {Pedri}}, \bibinfo {author}
  {\bibfnamefont {L.}~\bibnamefont {Vernac}}, \bibinfo {author} {\bibfnamefont
  {O.}~\bibnamefont {Gorceix}}, \ and\ \bibinfo {author} {\bibfnamefont
  {B.}~\bibnamefont {Laburthe-Tolra}},\ }\href {\doibase
  10.1103/PhysRevLett.105.040404} {\bibfield  {journal} {\bibinfo  {journal}
  {Phys. Rev. Lett.}\ }\textbf {\bibinfo {volume} {105}},\ \bibinfo {pages}
  {040404} (\bibinfo {year} {2010})}\BibitemShut {NoStop}%
\bibitem [{\citenamefont {Pasquiou}\ \emph {et~al.}(2011)\citenamefont
  {Pasquiou}, \citenamefont {Mar\'echal}, \citenamefont {Bismut}, \citenamefont
  {Pedri}, \citenamefont {Vernac}, \citenamefont {Gorceix},\ and\ \citenamefont
  {Laburthe-Tolra}}]{Pasquiou:2011}%
  \BibitemOpen
  \bibfield  {author} {\bibinfo {author} {\bibfnamefont {B.}~\bibnamefont
  {Pasquiou}}, \bibinfo {author} {\bibfnamefont {E.}~\bibnamefont
  {Mar\'echal}}, \bibinfo {author} {\bibfnamefont {G.}~\bibnamefont {Bismut}},
  \bibinfo {author} {\bibfnamefont {P.}~\bibnamefont {Pedri}}, \bibinfo
  {author} {\bibfnamefont {L.}~\bibnamefont {Vernac}}, \bibinfo {author}
  {\bibfnamefont {O.}~\bibnamefont {Gorceix}}, \ and\ \bibinfo {author}
  {\bibfnamefont {B.}~\bibnamefont {Laburthe-Tolra}},\ }\href {\doibase
  10.1103/PhysRevLett.106.255303} {\bibfield  {journal} {\bibinfo  {journal}
  {Phys. Rev. Lett.}\ }\textbf {\bibinfo {volume} {106}},\ \bibinfo {pages}
  {255303} (\bibinfo {year} {2011})}\BibitemShut {NoStop}%
\bibitem [{\citenamefont {Pasquiou}\ \emph {et~al.}(2012)\citenamefont
  {Pasquiou}, \citenamefont {Mar\'echal}, \citenamefont {Vernac}, \citenamefont
  {Gorceix},\ and\ \citenamefont {Laburthe-Tolra}}]{Pasquiou:2012}%
  \BibitemOpen
  \bibfield  {author} {\bibinfo {author} {\bibfnamefont {B.}~\bibnamefont
  {Pasquiou}}, \bibinfo {author} {\bibfnamefont {E.}~\bibnamefont
  {Mar\'echal}}, \bibinfo {author} {\bibfnamefont {L.}~\bibnamefont {Vernac}},
  \bibinfo {author} {\bibfnamefont {O.}~\bibnamefont {Gorceix}}, \ and\
  \bibinfo {author} {\bibfnamefont {B.}~\bibnamefont {Laburthe-Tolra}},\ }\href
  {\doibase 10.1103/PhysRevLett.108.045307} {\bibfield  {journal} {\bibinfo
  {journal} {Phys. Rev. Lett.}\ }\textbf {\bibinfo {volume} {108}},\ \bibinfo
  {pages} {045307} (\bibinfo {year} {2012})}\BibitemShut {NoStop}%
\bibitem [{\citenamefont {Bismut}\ \emph {et~al.}(2012)\citenamefont {Bismut},
  \citenamefont {Laburthe-Tolra}, \citenamefont {Mar\'echal}, \citenamefont
  {Pedri}, \citenamefont {Gorceix},\ and\ \citenamefont
  {Vernac}}]{Bismut:2012}%
  \BibitemOpen
  \bibfield  {author} {\bibinfo {author} {\bibfnamefont {G.}~\bibnamefont
  {Bismut}}, \bibinfo {author} {\bibfnamefont {B.}~\bibnamefont
  {Laburthe-Tolra}}, \bibinfo {author} {\bibfnamefont {E.}~\bibnamefont
  {Mar\'echal}}, \bibinfo {author} {\bibfnamefont {P.}~\bibnamefont {Pedri}},
  \bibinfo {author} {\bibfnamefont {O.}~\bibnamefont {Gorceix}}, \ and\
  \bibinfo {author} {\bibfnamefont {L.}~\bibnamefont {Vernac}},\ }\href
  {\doibase 10.1103/PhysRevLett.109.155302} {\bibfield  {journal} {\bibinfo
  {journal} {Phys. Rev. Lett.}\ }\textbf {\bibinfo {volume} {109}},\ \bibinfo
  {pages} {155302} (\bibinfo {year} {2012})}\BibitemShut {NoStop}%
\bibitem [{\citenamefont {de~Paz}\ \emph {et~al.}(2013)\citenamefont {de~Paz},
  \citenamefont {Sharma}, \citenamefont {Chotia}, \citenamefont {Mar\'echal},
  \citenamefont {Huckans}, \citenamefont {Pedri}, \citenamefont {Santos},
  \citenamefont {Gorceix}, \citenamefont {Vernac},\ and\ \citenamefont
  {Laburthe-Tolra}}]{dePaz:2013}%
  \BibitemOpen
  \bibfield  {author} {\bibinfo {author} {\bibfnamefont {A.}~\bibnamefont
  {de~Paz}}, \bibinfo {author} {\bibfnamefont {A.}~\bibnamefont {Sharma}},
  \bibinfo {author} {\bibfnamefont {A.}~\bibnamefont {Chotia}}, \bibinfo
  {author} {\bibfnamefont {E.}~\bibnamefont {Mar\'echal}}, \bibinfo {author}
  {\bibfnamefont {J.~H.}\ \bibnamefont {Huckans}}, \bibinfo {author}
  {\bibfnamefont {P.}~\bibnamefont {Pedri}}, \bibinfo {author} {\bibfnamefont
  {L.}~\bibnamefont {Santos}}, \bibinfo {author} {\bibfnamefont
  {O.}~\bibnamefont {Gorceix}}, \bibinfo {author} {\bibfnamefont
  {L.}~\bibnamefont {Vernac}}, \ and\ \bibinfo {author} {\bibfnamefont
  {B.}~\bibnamefont {Laburthe-Tolra}},\ }\href {\doibase
  10.1103/PhysRevLett.111.185305} {\bibfield  {journal} {\bibinfo  {journal}
  {Phys. Rev. Lett.}\ }\textbf {\bibinfo {volume} {111}},\ \bibinfo {pages}
  {185305} (\bibinfo {year} {2013})}\BibitemShut {NoStop}%
\bibitem [{\citenamefont {Aikawa}\ \emph
  {et~al.}(2014{\natexlab{a}})\citenamefont {Aikawa}, \citenamefont {Baier},
  \citenamefont {Frisch}, \citenamefont {Mark}, \citenamefont {Ravensbergen},\
  and\ \citenamefont {Ferlaino}}]{Aikawa:2014a}%
  \BibitemOpen
  \bibfield  {author} {\bibinfo {author} {\bibfnamefont {K.}~\bibnamefont
  {Aikawa}}, \bibinfo {author} {\bibfnamefont {S.}~\bibnamefont {Baier}},
  \bibinfo {author} {\bibfnamefont {A.}~\bibnamefont {Frisch}}, \bibinfo
  {author} {\bibfnamefont {M.}~\bibnamefont {Mark}}, \bibinfo {author}
  {\bibfnamefont {C.}~\bibnamefont {Ravensbergen}}, \ and\ \bibinfo {author}
  {\bibfnamefont {F.}~\bibnamefont {Ferlaino}},\ }\href {\doibase
  10.1126/science.1255259} {\bibfield  {journal} {\bibinfo  {journal}
  {Science}\ }\textbf {\bibinfo {volume} {345}},\ \bibinfo {pages} {1484}
  (\bibinfo {year} {2014}{\natexlab{a}})},\ \Eprint
  {http://arxiv.org/abs/http://science.sciencemag.org/content/345/6203/1484.full.pdf}
  {http://science.sciencemag.org/content/345/6203/1484.full.pdf} \BibitemShut
  {NoStop}%
\bibitem [{\citenamefont {Aikawa}\ \emph
  {et~al.}(2014{\natexlab{b}})\citenamefont {Aikawa}, \citenamefont {Frisch},
  \citenamefont {Mark}, \citenamefont {Baier}, \citenamefont {Grimm},
  \citenamefont {Bohn}, \citenamefont {Jin}, \citenamefont {Bruun},\ and\
  \citenamefont {Ferlaino}}]{Aikawa:2014b}%
  \BibitemOpen
  \bibfield  {author} {\bibinfo {author} {\bibfnamefont {K.}~\bibnamefont
  {Aikawa}}, \bibinfo {author} {\bibfnamefont {A.}~\bibnamefont {Frisch}},
  \bibinfo {author} {\bibfnamefont {M.}~\bibnamefont {Mark}}, \bibinfo {author}
  {\bibfnamefont {S.}~\bibnamefont {Baier}}, \bibinfo {author} {\bibfnamefont
  {R.}~\bibnamefont {Grimm}}, \bibinfo {author} {\bibfnamefont {J.~L.}\
  \bibnamefont {Bohn}}, \bibinfo {author} {\bibfnamefont {D.~S.}\ \bibnamefont
  {Jin}}, \bibinfo {author} {\bibfnamefont {G.~M.}\ \bibnamefont {Bruun}}, \
  and\ \bibinfo {author} {\bibfnamefont {F.}~\bibnamefont {Ferlaino}},\ }\href
  {\doibase 10.1103/PhysRevLett.113.263201} {\bibfield  {journal} {\bibinfo
  {journal} {Phys. Rev. Lett.}\ }\textbf {\bibinfo {volume} {113}},\ \bibinfo
  {pages} {263201} (\bibinfo {year} {2014}{\natexlab{b}})}\BibitemShut
  {NoStop}%
\bibitem [{\citenamefont {Burdick}\ \emph {et~al.}(2015)\citenamefont
  {Burdick}, \citenamefont {Baumann}, \citenamefont {Tang}, \citenamefont
  {Lu},\ and\ \citenamefont {Lev}}]{Burdick:2015}%
  \BibitemOpen
  \bibfield  {author} {\bibinfo {author} {\bibfnamefont {N.~Q.}\ \bibnamefont
  {Burdick}}, \bibinfo {author} {\bibfnamefont {K.}~\bibnamefont {Baumann}},
  \bibinfo {author} {\bibfnamefont {Y.}~\bibnamefont {Tang}}, \bibinfo {author}
  {\bibfnamefont {M.}~\bibnamefont {Lu}}, \ and\ \bibinfo {author}
  {\bibfnamefont {B.~L.}\ \bibnamefont {Lev}},\ }\href {\doibase
  10.1103/PhysRevLett.114.023201} {\bibfield  {journal} {\bibinfo  {journal}
  {Phys. Rev. Lett.}\ }\textbf {\bibinfo {volume} {114}},\ \bibinfo {pages}
  {023201} (\bibinfo {year} {2015})}\BibitemShut {NoStop}%
\bibitem [{\citenamefont {Baier}\ \emph {et~al.}(2016)\citenamefont {Baier},
  \citenamefont {Mark}, \citenamefont {Petter}, \citenamefont {Aikawa},
  \citenamefont {Chomaz}, \citenamefont {Cai}, \citenamefont {Baranov},
  \citenamefont {Zoller},\ and\ \citenamefont {Ferlaino}}]{Baier201}%
  \BibitemOpen
  \bibfield  {author} {\bibinfo {author} {\bibfnamefont {S.}~\bibnamefont
  {Baier}}, \bibinfo {author} {\bibfnamefont {M.~J.}\ \bibnamefont {Mark}},
  \bibinfo {author} {\bibfnamefont {D.}~\bibnamefont {Petter}}, \bibinfo
  {author} {\bibfnamefont {K.}~\bibnamefont {Aikawa}}, \bibinfo {author}
  {\bibfnamefont {L.}~\bibnamefont {Chomaz}}, \bibinfo {author} {\bibfnamefont
  {Z.}~\bibnamefont {Cai}}, \bibinfo {author} {\bibfnamefont {M.}~\bibnamefont
  {Baranov}}, \bibinfo {author} {\bibfnamefont {P.}~\bibnamefont {Zoller}}, \
  and\ \bibinfo {author} {\bibfnamefont {F.}~\bibnamefont {Ferlaino}},\ }\href
  {\doibase 10.1126/science.aac9812} {\bibfield  {journal} {\bibinfo  {journal}
  {Science}\ }\textbf {\bibinfo {volume} {352}},\ \bibinfo {pages} {201}
  (\bibinfo {year} {2016})},\ \Eprint
  {http://arxiv.org/abs/http://science.sciencemag.org/content/352/6282/201.full.pdf}
  {http://science.sciencemag.org/content/352/6282/201.full.pdf} \BibitemShut
  {NoStop}%
\bibitem [{\citenamefont {Tang}\ \emph {et~al.}(2016)\citenamefont {Tang},
  \citenamefont {Sykes}, \citenamefont {Burdick}, \citenamefont {DiSciacca},
  \citenamefont {Petrov},\ and\ \citenamefont {Lev}}]{Tang:2016}%
  \BibitemOpen
  \bibfield  {author} {\bibinfo {author} {\bibfnamefont {Y.}~\bibnamefont
  {Tang}}, \bibinfo {author} {\bibfnamefont {A.~G.}\ \bibnamefont {Sykes}},
  \bibinfo {author} {\bibfnamefont {N.~Q.}\ \bibnamefont {Burdick}}, \bibinfo
  {author} {\bibfnamefont {J.~M.}\ \bibnamefont {DiSciacca}}, \bibinfo {author}
  {\bibfnamefont {D.~S.}\ \bibnamefont {Petrov}}, \ and\ \bibinfo {author}
  {\bibfnamefont {B.~L.}\ \bibnamefont {Lev}},\ }\href@noop {} {\  (\bibinfo
  {year} {2016})},\ \Eprint {http://arxiv.org/abs/1606.08856} {1606.08856}
  \BibitemShut {NoStop}%
\bibitem [{\citenamefont {Rosensweig}(1985)}]{Rosensweig:1985}%
  \BibitemOpen
  \bibfield  {author} {\bibinfo {author} {\bibfnamefont {R.}~\bibnamefont
  {Rosensweig}},\ }\href@noop {} {\emph {\bibinfo {title}
  {Ferrohydrodynamics}}},\ Cambridge Monographs on Mechanics\ (\bibinfo
  {publisher} {Cambridge University Press},\ \bibinfo {year}
  {1985})\BibitemShut {NoStop}%
\bibitem [{Note1()}]{Note1}%
  \BibitemOpen
  \bibinfo {note} {While the model presented here, based on the two simple
  interaction potentials is sufficient to describe the experimental results
  observed so far, precise quantitative measurements might reveal new physics,
  in particular the interaction potential could take a more subtle
  form.}\BibitemShut {Stop}%
\bibitem [{\citenamefont {Santos}\ \emph {et~al.}(2003)\citenamefont {Santos},
  \citenamefont {Shlyapnikov},\ and\ \citenamefont {Lewenstein}}]{Santos:2003}%
  \BibitemOpen
  \bibfield  {author} {\bibinfo {author} {\bibfnamefont {L.}~\bibnamefont
  {Santos}}, \bibinfo {author} {\bibfnamefont {G.~V.}\ \bibnamefont
  {Shlyapnikov}}, \ and\ \bibinfo {author} {\bibfnamefont {M.}~\bibnamefont
  {Lewenstein}},\ }\href {\doibase 10.1103/PhysRevLett.90.250403} {\bibfield
  {journal} {\bibinfo  {journal} {Phys. Rev. Lett.}\ }\textbf {\bibinfo
  {volume} {90}},\ \bibinfo {pages} {250403} (\bibinfo {year}
  {2003})}\BibitemShut {NoStop}%
\bibitem [{\citenamefont {Wilson}\ \emph {et~al.}(2008)\citenamefont {Wilson},
  \citenamefont {Ronen}, \citenamefont {Bohn},\ and\ \citenamefont
  {Pu}}]{Wilson:2008}%
  \BibitemOpen
  \bibfield  {author} {\bibinfo {author} {\bibfnamefont {R.~M.}\ \bibnamefont
  {Wilson}}, \bibinfo {author} {\bibfnamefont {S.}~\bibnamefont {Ronen}},
  \bibinfo {author} {\bibfnamefont {J.~L.}\ \bibnamefont {Bohn}}, \ and\
  \bibinfo {author} {\bibfnamefont {H.}~\bibnamefont {Pu}},\ }\href {\doibase
  10.1103/PhysRevLett.100.245302} {\bibfield  {journal} {\bibinfo  {journal}
  {Phys. Rev. Lett.}\ }\textbf {\bibinfo {volume} {100}},\ \bibinfo {pages}
  {245302} (\bibinfo {year} {2008})}\BibitemShut {NoStop}%
\bibitem [{\citenamefont {Lahaye}\ \emph {et~al.}(2009)\citenamefont {Lahaye},
  \citenamefont {Menotti}, \citenamefont {Santos}, \citenamefont {Lewenstein},\
  and\ \citenamefont {Pfau}}]{Lahaye:2009}%
  \BibitemOpen
  \bibfield  {author} {\bibinfo {author} {\bibfnamefont {T.}~\bibnamefont
  {Lahaye}}, \bibinfo {author} {\bibfnamefont {C.}~\bibnamefont {Menotti}},
  \bibinfo {author} {\bibfnamefont {L.}~\bibnamefont {Santos}}, \bibinfo
  {author} {\bibfnamefont {M.}~\bibnamefont {Lewenstein}}, \ and\ \bibinfo
  {author} {\bibfnamefont {T.}~\bibnamefont {Pfau}},\ }\href
  {http://stacks.iop.org/0034-4885/72/i=12/a=126401} {\bibfield  {journal}
  {\bibinfo  {journal} {Reports on Progress in Physics}\ }\textbf {\bibinfo
  {volume} {72}},\ \bibinfo {pages} {126401} (\bibinfo {year}
  {2009})}\BibitemShut {NoStop}%
\bibitem [{\citenamefont {Wilson}\ \emph {et~al.}(2009)\citenamefont {Wilson},
  \citenamefont {Ronen},\ and\ \citenamefont {Bohn}}]{Wilson:2009a}%
  \BibitemOpen
  \bibfield  {author} {\bibinfo {author} {\bibfnamefont {R.~M.}\ \bibnamefont
  {Wilson}}, \bibinfo {author} {\bibfnamefont {S.}~\bibnamefont {Ronen}}, \
  and\ \bibinfo {author} {\bibfnamefont {J.~L.}\ \bibnamefont {Bohn}},\ }\href
  {\doibase 10.1103/PhysRevA.80.023614} {\bibfield  {journal} {\bibinfo
  {journal} {Phys. Rev. A}\ }\textbf {\bibinfo {volume} {80}},\ \bibinfo
  {pages} {023614} (\bibinfo {year} {2009})}\BibitemShut {NoStop}%
\bibitem [{\citenamefont {Wilson}\ \emph {et~al.}(2010)\citenamefont {Wilson},
  \citenamefont {Ronen},\ and\ \citenamefont {Bohn}}]{Wilson:2009b}%
  \BibitemOpen
  \bibfield  {author} {\bibinfo {author} {\bibfnamefont {R.~M.}\ \bibnamefont
  {Wilson}}, \bibinfo {author} {\bibfnamefont {S.}~\bibnamefont {Ronen}}, \
  and\ \bibinfo {author} {\bibfnamefont {J.~L.}\ \bibnamefont {Bohn}},\ }\href
  {\doibase 10.1103/PhysRevLett.104.094501} {\bibfield  {journal} {\bibinfo
  {journal} {Phys. Rev. Lett.}\ }\textbf {\bibinfo {volume} {104}},\ \bibinfo
  {pages} {094501} (\bibinfo {year} {2010})}\BibitemShut {NoStop}%
\bibitem [{\citenamefont {Nath}\ and\ \citenamefont
  {Santos}(2010)}]{Nath:2010}%
  \BibitemOpen
  \bibfield  {author} {\bibinfo {author} {\bibfnamefont {R.}~\bibnamefont
  {Nath}}\ and\ \bibinfo {author} {\bibfnamefont {L.}~\bibnamefont {Santos}},\
  }\href {\doibase 10.1103/PhysRevA.81.033626} {\bibfield  {journal} {\bibinfo
  {journal} {Phys. Rev. A}\ }\textbf {\bibinfo {volume} {81}},\ \bibinfo
  {pages} {033626} (\bibinfo {year} {2010})}\BibitemShut {NoStop}%
\bibitem [{\citenamefont {Ticknor}\ \emph {et~al.}(2011)\citenamefont
  {Ticknor}, \citenamefont {Wilson},\ and\ \citenamefont
  {Bohn}}]{Ticknor:2011}%
  \BibitemOpen
  \bibfield  {author} {\bibinfo {author} {\bibfnamefont {C.}~\bibnamefont
  {Ticknor}}, \bibinfo {author} {\bibfnamefont {R.~M.}\ \bibnamefont {Wilson}},
  \ and\ \bibinfo {author} {\bibfnamefont {J.~L.}\ \bibnamefont {Bohn}},\
  }\href {\doibase 10.1103/PhysRevLett.106.065301} {\bibfield  {journal}
  {\bibinfo  {journal} {Phys. Rev. Lett.}\ }\textbf {\bibinfo {volume} {106}},\
  \bibinfo {pages} {065301} (\bibinfo {year} {2011})}\BibitemShut {NoStop}%
\bibitem [{\citenamefont {Jona-Lasinio}\ \emph {et~al.}(2013)\citenamefont
  {Jona-Lasinio}, \citenamefont {\L{}akomy},\ and\ \citenamefont
  {Santos}}]{Jona-Lasinio:2013}%
  \BibitemOpen
  \bibfield  {author} {\bibinfo {author} {\bibfnamefont {M.}~\bibnamefont
  {Jona-Lasinio}}, \bibinfo {author} {\bibfnamefont {K.}~\bibnamefont
  {\L{}akomy}}, \ and\ \bibinfo {author} {\bibfnamefont {L.}~\bibnamefont
  {Santos}},\ }\href {\doibase 10.1103/PhysRevA.88.025603} {\bibfield
  {journal} {\bibinfo  {journal} {Phys. Rev. A}\ }\textbf {\bibinfo {volume}
  {88}},\ \bibinfo {pages} {025603} (\bibinfo {year} {2013})}\BibitemShut
  {NoStop}%
\bibitem [{\citenamefont {Bisset}\ and\ \citenamefont
  {Blakie}(2013)}]{Bisset:2013a}%
  \BibitemOpen
  \bibfield  {author} {\bibinfo {author} {\bibfnamefont {R.~N.}\ \bibnamefont
  {Bisset}}\ and\ \bibinfo {author} {\bibfnamefont {P.~B.}\ \bibnamefont
  {Blakie}},\ }\href {\doibase 10.1103/PhysRevLett.110.265302} {\bibfield
  {journal} {\bibinfo  {journal} {Phys. Rev. Lett.}\ }\textbf {\bibinfo
  {volume} {110}},\ \bibinfo {pages} {265302} (\bibinfo {year}
  {2013})}\BibitemShut {NoStop}%
\bibitem [{\citenamefont {Blakie}\ \emph {et~al.}(2013)\citenamefont {Blakie},
  \citenamefont {Baillie},\ and\ \citenamefont {Bisset}}]{Bisset:2013b}%
  \BibitemOpen
  \bibfield  {author} {\bibinfo {author} {\bibfnamefont {P.~B.}\ \bibnamefont
  {Blakie}}, \bibinfo {author} {\bibfnamefont {D.}~\bibnamefont {Baillie}}, \
  and\ \bibinfo {author} {\bibfnamefont {R.~N.}\ \bibnamefont {Bisset}},\
  }\href {\doibase 10.1103/PhysRevA.88.013638} {\bibfield  {journal} {\bibinfo
  {journal} {Phys. Rev. A}\ }\textbf {\bibinfo {volume} {88}},\ \bibinfo
  {pages} {013638} (\bibinfo {year} {2013})}\BibitemShut {NoStop}%
\bibitem [{\citenamefont {Ronen}\ \emph {et~al.}(2007)\citenamefont {Ronen},
  \citenamefont {Bortolotti},\ and\ \citenamefont {Bohn}}]{Ronen:2007}%
  \BibitemOpen
  \bibfield  {author} {\bibinfo {author} {\bibfnamefont {S.}~\bibnamefont
  {Ronen}}, \bibinfo {author} {\bibfnamefont {D.~C.~E.}\ \bibnamefont
  {Bortolotti}}, \ and\ \bibinfo {author} {\bibfnamefont {J.~L.}\ \bibnamefont
  {Bohn}},\ }\href {\doibase 10.1103/PhysRevLett.98.030406} {\bibfield
  {journal} {\bibinfo  {journal} {Phys. Rev. Lett.}\ }\textbf {\bibinfo
  {volume} {98}},\ \bibinfo {pages} {030406} (\bibinfo {year}
  {2007})}\BibitemShut {NoStop}%
\bibitem [{\citenamefont {Chin}\ \emph {et~al.}(2010)\citenamefont {Chin},
  \citenamefont {Grimm}, \citenamefont {Julienne},\ and\ \citenamefont
  {Tiesinga}}]{Chin:2008}%
  \BibitemOpen
  \bibfield  {author} {\bibinfo {author} {\bibfnamefont {C.}~\bibnamefont
  {Chin}}, \bibinfo {author} {\bibfnamefont {R.}~\bibnamefont {Grimm}},
  \bibinfo {author} {\bibfnamefont {P.}~\bibnamefont {Julienne}}, \ and\
  \bibinfo {author} {\bibfnamefont {E.}~\bibnamefont {Tiesinga}},\ }\href
  {\doibase 10.1103/RevModPhys.82.1225} {\bibfield  {journal} {\bibinfo
  {journal} {Rev. Mod. Phys.}\ }\textbf {\bibinfo {volume} {82}},\ \bibinfo
  {pages} {1225} (\bibinfo {year} {2010})}\BibitemShut {NoStop}%
\bibitem [{\citenamefont {Lu}\ \emph {et~al.}(2011)\citenamefont {Lu},
  \citenamefont {Burdick}, \citenamefont {Youn},\ and\ \citenamefont
  {Lev}}]{Lu:2011}%
  \BibitemOpen
  \bibfield  {author} {\bibinfo {author} {\bibfnamefont {M.}~\bibnamefont
  {Lu}}, \bibinfo {author} {\bibfnamefont {N.~Q.}\ \bibnamefont {Burdick}},
  \bibinfo {author} {\bibfnamefont {S.~H.}\ \bibnamefont {Youn}}, \ and\
  \bibinfo {author} {\bibfnamefont {B.~L.}\ \bibnamefont {Lev}},\ }\href
  {\doibase 10.1103/PhysRevLett.107.190401} {\bibfield  {journal} {\bibinfo
  {journal} {Phys. Rev. Lett.}\ }\textbf {\bibinfo {volume} {107}},\ \bibinfo
  {pages} {190401} (\bibinfo {year} {2011})}\BibitemShut {NoStop}%
\bibitem [{\citenamefont {Tang}\ \emph
  {et~al.}(2015{\natexlab{a}})\citenamefont {Tang}, \citenamefont {Burdick},
  \citenamefont {Baumann},\ and\ \citenamefont {Lev}}]{Tang:2015b}%
  \BibitemOpen
  \bibfield  {author} {\bibinfo {author} {\bibfnamefont {Y.}~\bibnamefont
  {Tang}}, \bibinfo {author} {\bibfnamefont {N.~Q.}\ \bibnamefont {Burdick}},
  \bibinfo {author} {\bibfnamefont {K.}~\bibnamefont {Baumann}}, \ and\
  \bibinfo {author} {\bibfnamefont {B.~L.}\ \bibnamefont {Lev}},\ }\href
  {http://stacks.iop.org/1367-2630/17/i=4/a=045006} {\bibfield  {journal}
  {\bibinfo  {journal} {New Journal of Physics}\ }\textbf {\bibinfo {volume}
  {17}},\ \bibinfo {pages} {045006} (\bibinfo {year}
  {2015}{\natexlab{a}})}\BibitemShut {NoStop}%
\bibitem [{\citenamefont {Aikawa}\ \emph {et~al.}(2012)\citenamefont {Aikawa},
  \citenamefont {Frisch}, \citenamefont {Mark}, \citenamefont {Baier},
  \citenamefont {Rietzler}, \citenamefont {Grimm},\ and\ \citenamefont
  {Ferlaino}}]{Aikawa:2012}%
  \BibitemOpen
  \bibfield  {author} {\bibinfo {author} {\bibfnamefont {K.}~\bibnamefont
  {Aikawa}}, \bibinfo {author} {\bibfnamefont {A.}~\bibnamefont {Frisch}},
  \bibinfo {author} {\bibfnamefont {M.}~\bibnamefont {Mark}}, \bibinfo {author}
  {\bibfnamefont {S.}~\bibnamefont {Baier}}, \bibinfo {author} {\bibfnamefont
  {A.}~\bibnamefont {Rietzler}}, \bibinfo {author} {\bibfnamefont
  {R.}~\bibnamefont {Grimm}}, \ and\ \bibinfo {author} {\bibfnamefont
  {F.}~\bibnamefont {Ferlaino}},\ }\href {\doibase
  10.1103/PhysRevLett.108.210401} {\bibfield  {journal} {\bibinfo  {journal}
  {Phys. Rev. Lett.}\ }\textbf {\bibinfo {volume} {108}},\ \bibinfo {pages}
  {210401} (\bibinfo {year} {2012})}\BibitemShut {NoStop}%
\bibitem [{\citenamefont {Werner}\ \emph {et~al.}(2005)\citenamefont {Werner},
  \citenamefont {Griesmaier}, \citenamefont {Hensler}, \citenamefont {Stuhler},
  \citenamefont {Pfau}, \citenamefont {Simoni},\ and\ \citenamefont
  {Tiesinga}}]{Werner:2005}%
  \BibitemOpen
  \bibfield  {author} {\bibinfo {author} {\bibfnamefont {J.}~\bibnamefont
  {Werner}}, \bibinfo {author} {\bibfnamefont {A.}~\bibnamefont {Griesmaier}},
  \bibinfo {author} {\bibfnamefont {S.}~\bibnamefont {Hensler}}, \bibinfo
  {author} {\bibfnamefont {J.}~\bibnamefont {Stuhler}}, \bibinfo {author}
  {\bibfnamefont {T.}~\bibnamefont {Pfau}}, \bibinfo {author} {\bibfnamefont
  {A.}~\bibnamefont {Simoni}}, \ and\ \bibinfo {author} {\bibfnamefont
  {E.}~\bibnamefont {Tiesinga}},\ }\href {\doibase
  10.1103/PhysRevLett.94.183201} {\bibfield  {journal} {\bibinfo  {journal}
  {Phys. Rev. Lett.}\ }\textbf {\bibinfo {volume} {94}},\ \bibinfo {pages}
  {183201} (\bibinfo {year} {2005})}\BibitemShut {NoStop}%
\bibitem [{\citenamefont {Frisch}\ \emph {et~al.}(2014)\citenamefont {Frisch},
  \citenamefont {Mark}, \citenamefont {Aikawa}, \citenamefont {Ferlaino},
  \citenamefont {Bohn}, \citenamefont {Makrides}, \citenamefont {Petrov},\ and\
  \citenamefont {Kotochigova}}]{Frisch:2014}%
  \BibitemOpen
  \bibfield  {author} {\bibinfo {author} {\bibfnamefont {A.}~\bibnamefont
  {Frisch}}, \bibinfo {author} {\bibfnamefont {M.}~\bibnamefont {Mark}},
  \bibinfo {author} {\bibfnamefont {K.}~\bibnamefont {Aikawa}}, \bibinfo
  {author} {\bibfnamefont {F.}~\bibnamefont {Ferlaino}}, \bibinfo {author}
  {\bibfnamefont {J.~L.}\ \bibnamefont {Bohn}}, \bibinfo {author}
  {\bibfnamefont {C.}~\bibnamefont {Makrides}}, \bibinfo {author}
  {\bibfnamefont {A.}~\bibnamefont {Petrov}}, \ and\ \bibinfo {author}
  {\bibfnamefont {S.}~\bibnamefont {Kotochigova}},\ }\href
  {http://dx.doi.org/10.1038/nature13137} {\bibfield  {journal} {\bibinfo
  {journal} {Nature}\ }\textbf {\bibinfo {volume} {507}},\ \bibinfo {pages}
  {475} (\bibinfo {year} {2014})}\BibitemShut {NoStop}%
\bibitem [{\citenamefont {Maier}\ \emph
  {et~al.}(2015{\natexlab{a}})\citenamefont {Maier}, \citenamefont
  {Ferrier-Barbut}, \citenamefont {Kadau}, \citenamefont {Schmitt},
  \citenamefont {Wenzel}, \citenamefont {Wink}, \citenamefont {Pfau},
  \citenamefont {Jachymski},\ and\ \citenamefont {Julienne}}]{Maier:2015a}%
  \BibitemOpen
  \bibfield  {author} {\bibinfo {author} {\bibfnamefont {T.}~\bibnamefont
  {Maier}}, \bibinfo {author} {\bibfnamefont {I.}~\bibnamefont
  {Ferrier-Barbut}}, \bibinfo {author} {\bibfnamefont {H.}~\bibnamefont
  {Kadau}}, \bibinfo {author} {\bibfnamefont {M.}~\bibnamefont {Schmitt}},
  \bibinfo {author} {\bibfnamefont {M.}~\bibnamefont {Wenzel}}, \bibinfo
  {author} {\bibfnamefont {C.}~\bibnamefont {Wink}}, \bibinfo {author}
  {\bibfnamefont {T.}~\bibnamefont {Pfau}}, \bibinfo {author} {\bibfnamefont
  {K.}~\bibnamefont {Jachymski}}, \ and\ \bibinfo {author} {\bibfnamefont
  {P.~S.}\ \bibnamefont {Julienne}},\ }\href {\doibase
  10.1103/PhysRevA.92.060702} {\bibfield  {journal} {\bibinfo  {journal} {Phys.
  Rev. A}\ }\textbf {\bibinfo {volume} {92}},\ \bibinfo {pages} {060702(R)}
  (\bibinfo {year} {2015}{\natexlab{a}})}\BibitemShut {NoStop}%
\bibitem [{\citenamefont {Baumann}\ \emph {et~al.}(2014)\citenamefont
  {Baumann}, \citenamefont {Burdick}, \citenamefont {Lu},\ and\ \citenamefont
  {Lev}}]{Baumann:2014}%
  \BibitemOpen
  \bibfield  {author} {\bibinfo {author} {\bibfnamefont {K.}~\bibnamefont
  {Baumann}}, \bibinfo {author} {\bibfnamefont {N.~Q.}\ \bibnamefont
  {Burdick}}, \bibinfo {author} {\bibfnamefont {M.}~\bibnamefont {Lu}}, \ and\
  \bibinfo {author} {\bibfnamefont {B.~L.}\ \bibnamefont {Lev}},\ }\href
  {\doibase 10.1103/PhysRevA.89.020701} {\bibfield  {journal} {\bibinfo
  {journal} {Phys. Rev. A}\ }\textbf {\bibinfo {volume} {89}},\ \bibinfo
  {pages} {020701} (\bibinfo {year} {2014})}\BibitemShut {NoStop}%
\bibitem [{\citenamefont {Maier}\ \emph
  {et~al.}(2015{\natexlab{b}})\citenamefont {Maier}, \citenamefont {Kadau},
  \citenamefont {Schmitt}, \citenamefont {Wenzel}, \citenamefont
  {Ferrier-Barbut}, \citenamefont {Pfau}, \citenamefont {Frisch}, \citenamefont
  {Baier}, \citenamefont {Aikawa}, \citenamefont {Chomaz}, \citenamefont
  {Mark}, \citenamefont {Ferlaino}, \citenamefont {Makrides}, \citenamefont
  {Tiesinga}, \citenamefont {Petrov},\ and\ \citenamefont
  {Kotochigova}}]{Maier:2015b}%
  \BibitemOpen
  \bibfield  {author} {\bibinfo {author} {\bibfnamefont {T.}~\bibnamefont
  {Maier}}, \bibinfo {author} {\bibfnamefont {H.}~\bibnamefont {Kadau}},
  \bibinfo {author} {\bibfnamefont {M.}~\bibnamefont {Schmitt}}, \bibinfo
  {author} {\bibfnamefont {M.}~\bibnamefont {Wenzel}}, \bibinfo {author}
  {\bibfnamefont {I.}~\bibnamefont {Ferrier-Barbut}}, \bibinfo {author}
  {\bibfnamefont {T.}~\bibnamefont {Pfau}}, \bibinfo {author} {\bibfnamefont
  {A.}~\bibnamefont {Frisch}}, \bibinfo {author} {\bibfnamefont
  {S.}~\bibnamefont {Baier}}, \bibinfo {author} {\bibfnamefont
  {K.}~\bibnamefont {Aikawa}}, \bibinfo {author} {\bibfnamefont
  {L.}~\bibnamefont {Chomaz}}, \bibinfo {author} {\bibfnamefont {M.~J.}\
  \bibnamefont {Mark}}, \bibinfo {author} {\bibfnamefont {F.}~\bibnamefont
  {Ferlaino}}, \bibinfo {author} {\bibfnamefont {C.}~\bibnamefont {Makrides}},
  \bibinfo {author} {\bibfnamefont {E.}~\bibnamefont {Tiesinga}}, \bibinfo
  {author} {\bibfnamefont {A.}~\bibnamefont {Petrov}}, \ and\ \bibinfo {author}
  {\bibfnamefont {S.}~\bibnamefont {Kotochigova}},\ }\href {\doibase
  10.1103/PhysRevX.5.041029} {\bibfield  {journal} {\bibinfo  {journal} {Phys.
  Rev. X}\ }\textbf {\bibinfo {volume} {5}},\ \bibinfo {pages} {041029}
  (\bibinfo {year} {2015}{\natexlab{b}})}\BibitemShut {NoStop}%
\bibitem [{\citenamefont {Frisch}\ \emph {et~al.}(2015)\citenamefont {Frisch},
  \citenamefont {Mark}, \citenamefont {Aikawa}, \citenamefont {Baier},
  \citenamefont {Grimm}, \citenamefont {Petrov}, \citenamefont {Kotochigova},
  \citenamefont {Qu\'em\'ener}, \citenamefont {Lepers}, \citenamefont
  {Dulieu},\ and\ \citenamefont {Ferlaino}}]{Frisch:2015}%
  \BibitemOpen
  \bibfield  {author} {\bibinfo {author} {\bibfnamefont {A.}~\bibnamefont
  {Frisch}}, \bibinfo {author} {\bibfnamefont {M.}~\bibnamefont {Mark}},
  \bibinfo {author} {\bibfnamefont {K.}~\bibnamefont {Aikawa}}, \bibinfo
  {author} {\bibfnamefont {S.}~\bibnamefont {Baier}}, \bibinfo {author}
  {\bibfnamefont {R.}~\bibnamefont {Grimm}}, \bibinfo {author} {\bibfnamefont
  {A.}~\bibnamefont {Petrov}}, \bibinfo {author} {\bibfnamefont
  {S.}~\bibnamefont {Kotochigova}}, \bibinfo {author} {\bibfnamefont
  {G.}~\bibnamefont {Qu\'em\'ener}}, \bibinfo {author} {\bibfnamefont
  {M.}~\bibnamefont {Lepers}}, \bibinfo {author} {\bibfnamefont
  {O.}~\bibnamefont {Dulieu}}, \ and\ \bibinfo {author} {\bibfnamefont
  {F.}~\bibnamefont {Ferlaino}},\ }\href {\doibase
  10.1103/PhysRevLett.115.203201} {\bibfield  {journal} {\bibinfo  {journal}
  {Phys. Rev. Lett.}\ }\textbf {\bibinfo {volume} {115}},\ \bibinfo {pages}
  {203201} (\bibinfo {year} {2015})}\BibitemShut {NoStop}%
\bibitem [{\citenamefont {Tang}\ \emph
  {et~al.}(2015{\natexlab{b}})\citenamefont {Tang}, \citenamefont {Sykes},
  \citenamefont {Burdick}, \citenamefont {Bohn},\ and\ \citenamefont
  {Lev}}]{Tang:2015a}%
  \BibitemOpen
  \bibfield  {author} {\bibinfo {author} {\bibfnamefont {Y.}~\bibnamefont
  {Tang}}, \bibinfo {author} {\bibfnamefont {A.}~\bibnamefont {Sykes}},
  \bibinfo {author} {\bibfnamefont {N.~Q.}\ \bibnamefont {Burdick}}, \bibinfo
  {author} {\bibfnamefont {J.~L.}\ \bibnamefont {Bohn}}, \ and\ \bibinfo
  {author} {\bibfnamefont {B.~L.}\ \bibnamefont {Lev}},\ }\href {\doibase
  10.1103/PhysRevA.92.022703} {\bibfield  {journal} {\bibinfo  {journal} {Phys.
  Rev. A}\ }\textbf {\bibinfo {volume} {92}},\ \bibinfo {pages} {022703}
  (\bibinfo {year} {2015}{\natexlab{b}})}\BibitemShut {NoStop}%
\bibitem [{\citenamefont {Kadau}\ \emph {et~al.}(2016)\citenamefont {Kadau},
  \citenamefont {Schmitt}, \citenamefont {Wenzel}, \citenamefont {Wink},
  \citenamefont {Maier}, \citenamefont {Ferrier-Barbut},\ and\ \citenamefont
  {Pfau}}]{Kadau:2016a}%
  \BibitemOpen
  \bibfield  {author} {\bibinfo {author} {\bibfnamefont {H.}~\bibnamefont
  {Kadau}}, \bibinfo {author} {\bibfnamefont {M.}~\bibnamefont {Schmitt}},
  \bibinfo {author} {\bibfnamefont {M.}~\bibnamefont {Wenzel}}, \bibinfo
  {author} {\bibfnamefont {C.}~\bibnamefont {Wink}}, \bibinfo {author}
  {\bibfnamefont {T.}~\bibnamefont {Maier}}, \bibinfo {author} {\bibfnamefont
  {I.}~\bibnamefont {Ferrier-Barbut}}, \ and\ \bibinfo {author} {\bibfnamefont
  {T.}~\bibnamefont {Pfau}},\ }\href {http://dx.doi.org/10.1038/nature16485}
  {\bibfield  {journal} {\bibinfo  {journal} {Nature}\ }\textbf {\bibinfo
  {volume} {530}},\ \bibinfo {pages} {194} (\bibinfo {year}
  {2016})}\BibitemShut {NoStop}%
\bibitem [{\citenamefont {Ferrier-Barbut}\ \emph {et~al.}(2016)\citenamefont
  {Ferrier-Barbut}, \citenamefont {Kadau}, \citenamefont {Schmitt},
  \citenamefont {Wenzel},\ and\ \citenamefont {Pfau}}]{Ferrier-Barbut:2016}%
  \BibitemOpen
  \bibfield  {author} {\bibinfo {author} {\bibfnamefont {I.}~\bibnamefont
  {Ferrier-Barbut}}, \bibinfo {author} {\bibfnamefont {H.}~\bibnamefont
  {Kadau}}, \bibinfo {author} {\bibfnamefont {M.}~\bibnamefont {Schmitt}},
  \bibinfo {author} {\bibfnamefont {M.}~\bibnamefont {Wenzel}}, \ and\ \bibinfo
  {author} {\bibfnamefont {T.}~\bibnamefont {Pfau}},\ }\href {\doibase
  10.1103/PhysRevLett.116.215301} {\bibfield  {journal} {\bibinfo  {journal}
  {Phys. Rev. Lett.}\ }\textbf {\bibinfo {volume} {116}},\ \bibinfo {pages}
  {215301} (\bibinfo {year} {2016})}\BibitemShut {NoStop}%
\bibitem [{Note2()}]{Note2}%
  \BibitemOpen
  \bibinfo {note} {The next order (nonlinearity) grows like
  $(l+2)(l+3)$}\BibitemShut {NoStop}%
\bibitem [{\citenamefont {Lima}\ and\ \citenamefont
  {Pelster}(2012)}]{Lima:2012}%
  \BibitemOpen
  \bibfield  {author} {\bibinfo {author} {\bibfnamefont {A.~R.~P.}\
  \bibnamefont {Lima}}\ and\ \bibinfo {author} {\bibfnamefont {A.}~\bibnamefont
  {Pelster}},\ }\href {\doibase 10.1103/PhysRevA.86.063609} {\bibfield
  {journal} {\bibinfo  {journal} {Phys. Rev. A}\ }\textbf {\bibinfo {volume}
  {86}},\ \bibinfo {pages} {063609} (\bibinfo {year} {2012})}\BibitemShut
  {NoStop}%
\bibitem [{\citenamefont {O'Dell}\ \emph {et~al.}(2004)\citenamefont {O'Dell},
  \citenamefont {Giovanazzi},\ and\ \citenamefont {Eberlein}}]{ODell:2004}%
  \BibitemOpen
  \bibfield  {author} {\bibinfo {author} {\bibfnamefont {D.~H.~J.}\
  \bibnamefont {O'Dell}}, \bibinfo {author} {\bibfnamefont {S.}~\bibnamefont
  {Giovanazzi}}, \ and\ \bibinfo {author} {\bibfnamefont {C.}~\bibnamefont
  {Eberlein}},\ }\href {\doibase 10.1103/PhysRevLett.92.250401} {\bibfield
  {journal} {\bibinfo  {journal} {Phys. Rev. Lett.}\ }\textbf {\bibinfo
  {volume} {92}},\ \bibinfo {pages} {250401} (\bibinfo {year}
  {2004})}\BibitemShut {NoStop}%
\bibitem [{\citenamefont {Eberlein}\ \emph {et~al.}(2005)\citenamefont
  {Eberlein}, \citenamefont {Giovanazzi},\ and\ \citenamefont
  {O'Dell}}]{Eberlein:2005}%
  \BibitemOpen
  \bibfield  {author} {\bibinfo {author} {\bibfnamefont {C.}~\bibnamefont
  {Eberlein}}, \bibinfo {author} {\bibfnamefont {S.}~\bibnamefont
  {Giovanazzi}}, \ and\ \bibinfo {author} {\bibfnamefont {D.~H.~J.}\
  \bibnamefont {O'Dell}},\ }\href {\doibase 10.1103/PhysRevA.71.033618}
  {\bibfield  {journal} {\bibinfo  {journal} {Phys. Rev. A}\ }\textbf {\bibinfo
  {volume} {71}},\ \bibinfo {pages} {033618} (\bibinfo {year}
  {2005})}\BibitemShut {NoStop}%
\bibitem [{\citenamefont {Bulgac}(2002)}]{Bulgac:2002}%
  \BibitemOpen
  \bibfield  {author} {\bibinfo {author} {\bibfnamefont {A.}~\bibnamefont
  {Bulgac}},\ }\href {\doibase 10.1103/PhysRevLett.89.050402} {\bibfield
  {journal} {\bibinfo  {journal} {Phys. Rev. Lett.}\ }\textbf {\bibinfo
  {volume} {89}},\ \bibinfo {pages} {050402} (\bibinfo {year}
  {2002})}\BibitemShut {NoStop}%
\bibitem [{\citenamefont {Xi}\ and\ \citenamefont {Saito}(2015)}]{Xi:2015}%
  \BibitemOpen
  \bibfield  {author} {\bibinfo {author} {\bibfnamefont {K.-T.}\ \bibnamefont
  {Xi}}\ and\ \bibinfo {author} {\bibfnamefont {H.}~\bibnamefont {Saito}},\
  }\href@noop {} {\  (\bibinfo {year} {2015})},\ \Eprint
  {http://arxiv.org/abs/1510.07842} {arXiv:1510.07842} \BibitemShut {NoStop}%
\bibitem [{\citenamefont {Bisset}\ and\ \citenamefont
  {Blakie}(2015)}]{Bisset:2015}%
  \BibitemOpen
  \bibfield  {author} {\bibinfo {author} {\bibfnamefont {R.~N.}\ \bibnamefont
  {Bisset}}\ and\ \bibinfo {author} {\bibfnamefont {P.~B.}\ \bibnamefont
  {Blakie}},\ }\href {\doibase 10.1103/PhysRevA.92.061603} {\bibfield
  {journal} {\bibinfo  {journal} {Phys. Rev. A}\ }\textbf {\bibinfo {volume}
  {92}},\ \bibinfo {pages} {061603} (\bibinfo {year} {2015})}\BibitemShut
  {NoStop}%
\bibitem [{\citenamefont {Blakie}(2016)}]{Blakie:2016}%
  \BibitemOpen
  \bibfield  {author} {\bibinfo {author} {\bibfnamefont {P.~B.}\ \bibnamefont
  {Blakie}},\ }\href {\doibase 10.1103/PhysRevA.93.033644} {\bibfield
  {journal} {\bibinfo  {journal} {Phys. Rev. A}\ }\textbf {\bibinfo {volume}
  {93}},\ \bibinfo {pages} {033644} (\bibinfo {year} {2016})}\BibitemShut
  {NoStop}%
\bibitem [{\citenamefont {Petrov}(2015)}]{Petrov:2015}%
  \BibitemOpen
  \bibfield  {author} {\bibinfo {author} {\bibfnamefont {D.~S.}\ \bibnamefont
  {Petrov}},\ }\href {\doibase 10.1103/PhysRevLett.115.155302} {\bibfield
  {journal} {\bibinfo  {journal} {Phys. Rev. Lett.}\ }\textbf {\bibinfo
  {volume} {115}},\ \bibinfo {pages} {155302} (\bibinfo {year}
  {2015})}\BibitemShut {NoStop}%
\bibitem [{\citenamefont {Volovik}(2009)}]{Volovik:2009}%
  \BibitemOpen
  \bibfield  {author} {\bibinfo {author} {\bibfnamefont {G.~E.}\ \bibnamefont
  {Volovik}},\ }\href {\doibase 10.1093/acprof:oso/9780199564842.001.0001}
  {\emph {\bibinfo {title} {{The Universe in a Helium Droplet}}}},\
  International Series of Monographs on Physics\ (\bibinfo  {publisher} {Oxford
  University Press},\ \bibinfo {address} {Oxford},\ \bibinfo {year}
  {2009})\BibitemShut {NoStop}%
\bibitem [{\citenamefont {Lee}\ and\ \citenamefont {Yang}(1957)}]{Lee:1957a}%
  \BibitemOpen
  \bibfield  {author} {\bibinfo {author} {\bibfnamefont {T.~D.}\ \bibnamefont
  {Lee}}\ and\ \bibinfo {author} {\bibfnamefont {C.~N.}\ \bibnamefont {Yang}},\
  }\href {\doibase 10.1103/PhysRev.105.1119} {\bibfield  {journal} {\bibinfo
  {journal} {Phys. Rev.}\ }\textbf {\bibinfo {volume} {105}},\ \bibinfo {pages}
  {1119} (\bibinfo {year} {1957})}\BibitemShut {NoStop}%
\bibitem [{\citenamefont {Lee}\ \emph {et~al.}(1957)\citenamefont {Lee},
  \citenamefont {Huang},\ and\ \citenamefont {Yang}}]{Lee:1957b}%
  \BibitemOpen
  \bibfield  {author} {\bibinfo {author} {\bibfnamefont {T.~D.}\ \bibnamefont
  {Lee}}, \bibinfo {author} {\bibfnamefont {K.}~\bibnamefont {Huang}}, \ and\
  \bibinfo {author} {\bibfnamefont {C.~N.}\ \bibnamefont {Yang}},\ }\href
  {\doibase 10.1103/PhysRev.106.1135} {\bibfield  {journal} {\bibinfo
  {journal} {Phys. Rev.}\ }\textbf {\bibinfo {volume} {106}},\ \bibinfo {pages}
  {1135} (\bibinfo {year} {1957})}\BibitemShut {NoStop}%
\bibitem [{\citenamefont {Lima}\ and\ \citenamefont
  {Pelster}(2011)}]{Lima:2011}%
  \BibitemOpen
  \bibfield  {author} {\bibinfo {author} {\bibfnamefont {A.~R.~P.}\
  \bibnamefont {Lima}}\ and\ \bibinfo {author} {\bibfnamefont {A.}~\bibnamefont
  {Pelster}},\ }\href {\doibase 10.1103/PhysRevA.84.041604} {\bibfield
  {journal} {\bibinfo  {journal} {Phys. Rev. A}\ }\textbf {\bibinfo {volume}
  {84}},\ \bibinfo {pages} {041604} (\bibinfo {year} {2011})}\BibitemShut
  {NoStop}%
\bibitem [{\citenamefont {Sch\"utzhold}\ \emph {et~al.}(2006)\citenamefont
  {Sch\"utzhold}, \citenamefont {Uhlmann}, \citenamefont {Xu},\ and\
  \citenamefont {Fischer}}]{Scutzhold:2006}%
  \BibitemOpen
  \bibfield  {author} {\bibinfo {author} {\bibfnamefont {R.}~\bibnamefont
  {Sch\"utzhold}}, \bibinfo {author} {\bibfnamefont {M.}~\bibnamefont
  {Uhlmann}}, \bibinfo {author} {\bibfnamefont {Y.}~\bibnamefont {Xu}}, \ and\
  \bibinfo {author} {\bibfnamefont {U.~R.}\ \bibnamefont {Fischer}},\ }\href
  {\doibase 10.1142/S0217979206035631} {\bibfield  {journal} {\bibinfo
  {journal} {International Journal of Modern Physics B}\ }\textbf {\bibinfo
  {volume} {20}},\ \bibinfo {pages} {3555} (\bibinfo {year}
  {2006})}\BibitemShut {NoStop}%
\bibitem [{\citenamefont {W\"achtler}\ and\ \citenamefont
  {Santos}(2016)}]{Wachtler:2016a}%
  \BibitemOpen
  \bibfield  {author} {\bibinfo {author} {\bibfnamefont {F.}~\bibnamefont
  {W\"achtler}}\ and\ \bibinfo {author} {\bibfnamefont {L.}~\bibnamefont
  {Santos}},\ }\href {\doibase 10.1103/PhysRevA.93.061603} {\bibfield
  {journal} {\bibinfo  {journal} {Phys. Rev. A}\ }\textbf {\bibinfo {volume}
  {93}},\ \bibinfo {pages} {061603} (\bibinfo {year} {2016})}\BibitemShut
  {NoStop}%
\bibitem [{\citenamefont {W{\"a}chtler}\ and\ \citenamefont
  {Santos}(2016)}]{Wachtler:2016b}%
  \BibitemOpen
  \bibfield  {author} {\bibinfo {author} {\bibfnamefont {F.}~\bibnamefont
  {W{\"a}chtler}}\ and\ \bibinfo {author} {\bibfnamefont {L.}~\bibnamefont
  {Santos}},\ }\href@noop {} {\  (\bibinfo {year} {2016})},\ \Eprint
  {http://arxiv.org/abs/1605.08676} {1605.08676} \BibitemShut {NoStop}%
\bibitem [{\citenamefont {Bisset}\ \emph {et~al.}(2016)\citenamefont {Bisset},
  \citenamefont {Wilson}, \citenamefont {Baillie},\ and\ \citenamefont
  {Blakie}}]{Bisset:2016}%
  \BibitemOpen
  \bibfield  {author} {\bibinfo {author} {\bibfnamefont {R.~N.}\ \bibnamefont
  {Bisset}}, \bibinfo {author} {\bibfnamefont {R.~M.}\ \bibnamefont {Wilson}},
  \bibinfo {author} {\bibfnamefont {D.}~\bibnamefont {Baillie}}, \ and\
  \bibinfo {author} {\bibfnamefont {P.~B.}\ \bibnamefont {Blakie}},\
  }\href@noop {} {\  (\bibinfo {year} {2016})},\ \Eprint
  {http://arxiv.org/abs/1605.04964} {1605.04964} \BibitemShut {NoStop}%
\bibitem [{\citenamefont {Saito}(2016)}]{Saito:2016}%
  \BibitemOpen
  \bibfield  {author} {\bibinfo {author} {\bibfnamefont {H.}~\bibnamefont
  {Saito}},\ }\href {\doibase 10.7566/JPSJ.85.053001} {\bibfield  {journal}
  {\bibinfo  {journal} {Journal of the Physical Society of Japan}\ }\textbf
  {\bibinfo {volume} {85}},\ \bibinfo {pages} {053001} (\bibinfo {year}
  {2016})},\ \Eprint
  {http://arxiv.org/abs/http://dx.doi.org/10.7566/JPSJ.85.053001}
  {http://dx.doi.org/10.7566/JPSJ.85.053001} \BibitemShut {NoStop}%
\bibitem [{\citenamefont {Kadau}(2016)}]{Kadau:2016b}%
  \BibitemOpen
  \bibfield  {author} {\bibinfo {author} {\bibfnamefont {H.}~\bibnamefont
  {Kadau}},\ }\href@noop {} {Ph.D. thesis},\ \bibinfo  {school} {Stuttgart
  University} (\bibinfo {year} {2016})\BibitemShut {NoStop}%
\bibitem [{\citenamefont {Schmitt}\ \emph {et~al.}(2016)\citenamefont
  {Schmitt}, \citenamefont {Wenzel}, \citenamefont {B{\"o}ttcher},
  \citenamefont {Ferrier-Barbut},\ and\ \citenamefont {Pfau}}]{Schmitt:2016}%
  \BibitemOpen
  \bibfield  {author} {\bibinfo {author} {\bibfnamefont {M.}~\bibnamefont
  {Schmitt}}, \bibinfo {author} {\bibfnamefont {M.}~\bibnamefont {Wenzel}},
  \bibinfo {author} {\bibfnamefont {F.}~\bibnamefont {B{\"o}ttcher}}, \bibinfo
  {author} {\bibfnamefont {I.}~\bibnamefont {Ferrier-Barbut}}, \ and\ \bibinfo
  {author} {\bibfnamefont {T.}~\bibnamefont {Pfau}},\ }\href@noop {} {\bibfield
   {journal} {\bibinfo  {journal} {ArXiv.org}\ ,\ \bibinfo {pages}
  {arXiv:1607.07355}} (\bibinfo {year} {2016})},\ \Eprint
  {http://arxiv.org/abs/1607.07355} {1607.07355} \BibitemShut {NoStop}%
\bibitem [{\citenamefont {Baillie}\ \emph {et~al.}(2016)\citenamefont
  {Baillie}, \citenamefont {Wilson}, \citenamefont {Bisset},\ and\
  \citenamefont {Blakie}}]{Baillie:2016}%
  \BibitemOpen
  \bibfield  {author} {\bibinfo {author} {\bibfnamefont {D.}~\bibnamefont
  {Baillie}}, \bibinfo {author} {\bibfnamefont {R.~M.}\ \bibnamefont {Wilson}},
  \bibinfo {author} {\bibfnamefont {R.~N.}\ \bibnamefont {Bisset}}, \ and\
  \bibinfo {author} {\bibfnamefont {P.~B.}\ \bibnamefont {Blakie}},\
  }\href@noop {} {\  (\bibinfo {year} {2016})},\ \Eprint
  {http://arxiv.org/abs/1606.00824} {1606.00824} \BibitemShut {NoStop}%
\bibitem [{\citenamefont {Chomaz}\ \emph {et~al.}(2016)\citenamefont {Chomaz},
  \citenamefont {Baier}, \citenamefont {Petter}, \citenamefont {Mark},
  \citenamefont {W{\"a}chtler}, \citenamefont {Santos},\ and\ \citenamefont
  {Ferlaino}}]{Chomaz:2016}%
  \BibitemOpen
  \bibfield  {author} {\bibinfo {author} {\bibfnamefont {L.}~\bibnamefont
  {Chomaz}}, \bibinfo {author} {\bibfnamefont {S.}~\bibnamefont {Baier}},
  \bibinfo {author} {\bibfnamefont {D.}~\bibnamefont {Petter}}, \bibinfo
  {author} {\bibfnamefont {M.~J.}\ \bibnamefont {Mark}}, \bibinfo {author}
  {\bibfnamefont {F.}~\bibnamefont {W{\"a}chtler}}, \bibinfo {author}
  {\bibfnamefont {L.}~\bibnamefont {Santos}}, \ and\ \bibinfo {author}
  {\bibfnamefont {F.}~\bibnamefont {Ferlaino}},\ }\href@noop {} {\bibfield
  {journal} {\bibinfo  {journal} {ArXiv.org}\ ,\ \bibinfo {pages}
  {arXiv:1607.06613}} (\bibinfo {year} {2016})},\ \Eprint
  {http://arxiv.org/abs/1607.06613} {1607.06613} \BibitemShut {NoStop}%
\bibitem [{\citenamefont {Dalfovo}\ and\ \citenamefont
  {Stringari}(2001)}]{Dalfovo:2001}%
  \BibitemOpen
  \bibfield  {author} {\bibinfo {author} {\bibfnamefont {F.}~\bibnamefont
  {Dalfovo}}\ and\ \bibinfo {author} {\bibfnamefont {S.}~\bibnamefont
  {Stringari}},\ }\href {\doibase http://dx.doi.org/10.1063/1.1424926}
  {\bibfield  {journal} {\bibinfo  {journal} {The Journal of Chemical Physics}\
  }\textbf {\bibinfo {volume} {115}},\ \bibinfo {pages} {10078} (\bibinfo
  {year} {2001})}\BibitemShut {NoStop}%
\bibitem [{\citenamefont {Casas}\ and\ \citenamefont
  {Stringari}(1990)}]{Casas:1990}%
  \BibitemOpen
  \bibfield  {author} {\bibinfo {author} {\bibfnamefont {M.}~\bibnamefont
  {Casas}}\ and\ \bibinfo {author} {\bibfnamefont {S.}~\bibnamefont
  {Stringari}},\ }\href {\doibase 10.1007/BF00692450} {\bibfield  {journal}
  {\bibinfo  {journal} {Journal of Low Temperature Physics}\ }\textbf {\bibinfo
  {volume} {79}},\ \bibinfo {pages} {135} (\bibinfo {year} {1990})}\BibitemShut
  {NoStop}%
\bibitem [{\citenamefont {Gomez}\ \emph {et~al.}(2014)\citenamefont {Gomez},
  \citenamefont {Ferguson}, \citenamefont {Cryan}, \citenamefont {Bacellar},
  \citenamefont {Tanyag}, \citenamefont {Jones}, \citenamefont {Schorb},
  \citenamefont {Anielski}, \citenamefont {Belkacem}, \citenamefont {Bernando},
  \citenamefont {Boll}, \citenamefont {Bozek}, \citenamefont {Carron},
  \citenamefont {Chen}, \citenamefont {Delmas}, \citenamefont {Englert},
  \citenamefont {Epp}, \citenamefont {Erk}, \citenamefont {Foucar},
  \citenamefont {Hartmann}, \citenamefont {Hexemer}, \citenamefont {Huth},
  \citenamefont {Kwok}, \citenamefont {Leone}, \citenamefont {Ma},
  \citenamefont {Maia}, \citenamefont {Malmerberg}, \citenamefont {Marchesini},
  \citenamefont {Neumark}, \citenamefont {Poon}, \citenamefont {Prell},
  \citenamefont {Rolles}, \citenamefont {Rudek}, \citenamefont {Rudenko},
  \citenamefont {Seifrid}, \citenamefont {Siefermann}, \citenamefont {Sturm},
  \citenamefont {Swiggers}, \citenamefont {Ullrich}, \citenamefont {Weise},
  \citenamefont {Zwart}, \citenamefont {Bostedt}, \citenamefont {Gessner},\
  and\ \citenamefont {Vilesov}}]{Gomez:2014}%
  \BibitemOpen
  \bibfield  {author} {\bibinfo {author} {\bibfnamefont {L.~F.}\ \bibnamefont
  {Gomez}}, \bibinfo {author} {\bibfnamefont {K.~R.}\ \bibnamefont {Ferguson}},
  \bibinfo {author} {\bibfnamefont {J.~P.}\ \bibnamefont {Cryan}}, \bibinfo
  {author} {\bibfnamefont {C.}~\bibnamefont {Bacellar}}, \bibinfo {author}
  {\bibfnamefont {R.~M.~P.}\ \bibnamefont {Tanyag}}, \bibinfo {author}
  {\bibfnamefont {C.}~\bibnamefont {Jones}}, \bibinfo {author} {\bibfnamefont
  {S.}~\bibnamefont {Schorb}}, \bibinfo {author} {\bibfnamefont
  {D.}~\bibnamefont {Anielski}}, \bibinfo {author} {\bibfnamefont
  {A.}~\bibnamefont {Belkacem}}, \bibinfo {author} {\bibfnamefont
  {C.}~\bibnamefont {Bernando}}, \bibinfo {author} {\bibfnamefont
  {R.}~\bibnamefont {Boll}}, \bibinfo {author} {\bibfnamefont {J.}~\bibnamefont
  {Bozek}}, \bibinfo {author} {\bibfnamefont {S.}~\bibnamefont {Carron}},
  \bibinfo {author} {\bibfnamefont {G.}~\bibnamefont {Chen}}, \bibinfo {author}
  {\bibfnamefont {T.}~\bibnamefont {Delmas}}, \bibinfo {author} {\bibfnamefont
  {L.}~\bibnamefont {Englert}}, \bibinfo {author} {\bibfnamefont {S.~W.}\
  \bibnamefont {Epp}}, \bibinfo {author} {\bibfnamefont {B.}~\bibnamefont
  {Erk}}, \bibinfo {author} {\bibfnamefont {L.}~\bibnamefont {Foucar}},
  \bibinfo {author} {\bibfnamefont {R.}~\bibnamefont {Hartmann}}, \bibinfo
  {author} {\bibfnamefont {A.}~\bibnamefont {Hexemer}}, \bibinfo {author}
  {\bibfnamefont {M.}~\bibnamefont {Huth}}, \bibinfo {author} {\bibfnamefont
  {J.}~\bibnamefont {Kwok}}, \bibinfo {author} {\bibfnamefont {S.~R.}\
  \bibnamefont {Leone}}, \bibinfo {author} {\bibfnamefont {J.~H.~S.}\
  \bibnamefont {Ma}}, \bibinfo {author} {\bibfnamefont {F.~R. N.~C.}\
  \bibnamefont {Maia}}, \bibinfo {author} {\bibfnamefont {E.}~\bibnamefont
  {Malmerberg}}, \bibinfo {author} {\bibfnamefont {S.}~\bibnamefont
  {Marchesini}}, \bibinfo {author} {\bibfnamefont {D.~M.}\ \bibnamefont
  {Neumark}}, \bibinfo {author} {\bibfnamefont {B.}~\bibnamefont {Poon}},
  \bibinfo {author} {\bibfnamefont {J.}~\bibnamefont {Prell}}, \bibinfo
  {author} {\bibfnamefont {D.}~\bibnamefont {Rolles}}, \bibinfo {author}
  {\bibfnamefont {B.}~\bibnamefont {Rudek}}, \bibinfo {author} {\bibfnamefont
  {A.}~\bibnamefont {Rudenko}}, \bibinfo {author} {\bibfnamefont
  {M.}~\bibnamefont {Seifrid}}, \bibinfo {author} {\bibfnamefont {K.~R.}\
  \bibnamefont {Siefermann}}, \bibinfo {author} {\bibfnamefont {F.~P.}\
  \bibnamefont {Sturm}}, \bibinfo {author} {\bibfnamefont {M.}~\bibnamefont
  {Swiggers}}, \bibinfo {author} {\bibfnamefont {J.}~\bibnamefont {Ullrich}},
  \bibinfo {author} {\bibfnamefont {F.}~\bibnamefont {Weise}}, \bibinfo
  {author} {\bibfnamefont {P.}~\bibnamefont {Zwart}}, \bibinfo {author}
  {\bibfnamefont {C.}~\bibnamefont {Bostedt}}, \bibinfo {author} {\bibfnamefont
  {O.}~\bibnamefont {Gessner}}, \ and\ \bibinfo {author} {\bibfnamefont
  {A.~F.}\ \bibnamefont {Vilesov}},\ }\href {\doibase 10.1126/science.1252395}
  {\bibfield  {journal} {\bibinfo  {journal} {Science}\ }\textbf {\bibinfo
  {volume} {345}},\ \bibinfo {pages} {906} (\bibinfo {year} {2014})},\ \Eprint
  {http://arxiv.org/abs/http://science.sciencemag.org/content/345/6199/906.full.pdf}
  {http://science.sciencemag.org/content/345/6199/906.full.pdf} \BibitemShut
  {NoStop}%
\end{thebibliography}

\end{document}